\documentclass[sigplan, screen]{acmart}
\AtBeginDocument{%
  \providecommand\BibTeX{{%
    \normalfont B\kern-0.5em{\scshape i\kern-0.25em b}\kern-0.8em\TeX}}}

\copyrightyear{2025}
\acmYear{2025}
\setcopyright{rightsretained}
\acmConference[ASPLOS '25]{Proceedings of the 30th ACM International Conference on Architectural Support for Programming Languages and Operating Systems, Volume 1}{March 30--April 3, 2025}{Rotterdam, Netherlands}

\acmBooktitle{Proceedings of the 30th ACM International Conference on Architectural Support for Programming Languages and Operating Systems, Volume 1 (ASPLOS '25), March 30--April 3, 2025, Rotterdam, Netherlands}
\acmDOI{10.1145/3669940.3707251}
\acmISBN{979-8-4007-0698-1/25/03}
\makeatletter
\gdef\@copyrightpermission{
  \begin{minipage}{0.3\columnwidth}
   \href{https://creativecommons.org/licenses/by-nc-sa/4.0/}{\includegraphics[width=0.90\textwidth]{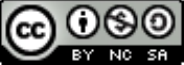}}
  \end{minipage}\hfill
  \begin{minipage}{0.7\columnwidth}
   \href{https://creativecommons.org/licenses/by-nc-sa/4.0/}{\textcolor{black}{This work is licensed under a Creative Commons Attribution-NonCommercial-ShareAlike International 4.0 License.}}
  \end{minipage}
  \vspace{5pt}
}
\makeatother

\usepackage{algorithm}
\usepackage{graphicx}
\usepackage{algpseudocode}
\usepackage{amsmath}
\usepackage{subfigure}
\usepackage{subcaption}
\usepackage{caption}
\usepackage{ulem} % lcc 
\usepackage{booktabs}
\usepackage{tabularx}
\usepackage{xcolor} 
\usepackage{enumitem}
\usepackage{multirow} 
\usepackage{calc}
\usepackage{pifont}
\usepackage{amsthm}
\usepackage{orcidlink}
\usepackage[]{hyperref}
\hypersetup{
    colorlinks=true,       
    linkcolor=blue,       
    citecolor=blue,       
    urlcolor=blue        
}

\makeatletter
\renewcommand{\ALG@name}{Algorithm}

\algrenewcommand\alglinenumber[1]{\tiny #1:}
\algrenewcommand\algorithmicindent{1.0em}%
\makeatother

\algdef{SE}[SWITCH]{Switch}{EndSwitch}[1]{\textbf{switch} #1\ \algorithmicdo}{\algorithmicend\ \textbf{switch}}%
\algdef{SE}[CASE]{Case}{EndCase}[1]{\textbf{case} #1:}{\algorithmicend\ \textbf{case}}%
\algtext*{EndSwitch}%
\algtext*{EndCase}%
\algtext*{EndProcedure} % Hide end procedure
\algtext*{EndIf} % Hide end procedure
\algtext*{EndFor} % Hide end procedure
\algtext*{EndWhile}
\algtext*{EndFunction}

\begin{document}
\makeatletter
\def\@copyrightspace{\relax}
\makeatother
\title{\projectname{}: Enabling GPU Resourcing-on-Demand for Serverless DL Serving via Introspective Elasticity}
\author{Cunchi Lv}
\email{lvcunchi21s@ict.ac.cn}
\affiliation{
  \institution{ICT, CAS}
  \institution{UCAS}
  \institution{Zhongguancun Lab}
  \country{Beijing, China}
  % \location{Beijing, China}
}

% \author{\href{https://orcid.org/0000-0001-7105-8355}{Xiao Shi}}
\author{Xiao Shi}
\email{shixiao@ict.ac.cn}
\authornote{Corresponding author.}
\affiliation{
  \institution{ICT, CAS}
  \institution{Nanjing Institute of InforSuperBahn}
  \country{Beijing, China}
  % \location{Beijing, China}
}

% \author{\href{https://orcid.org/0009-0004-2170-6560}{Zhengyu Lei}}
\author{Zhengyu Lei}
\email{leizhengyu20s@ict.ac.cn}
\affiliation{
  \institution{ICT, CAS}
  \institution{UCAS}
  \institution{Zhongguancun Lab}
  \country{Beijing, China}
  % \location{Beijing, China}
}

% \author{\href{https://orcid.org/0009-0004-9015-777X}{Jinyue Huang}}
\author{Jinyue Huang}
\email{huangjinyue22s@ict.ac.cn}
\affiliation{
  \institution{ICT, CAS}
  \institution{UCAS}
  \country{Beijing, China}
  % \location{Beijing, China}
}

% \vspace{-0.1pt}

\author{Wenting Tan}
\email{tanwenting@ict.ac.cn}
\affiliation{
  \institution{ICT, CAS}
  \country{Beijing, China}
  % \location{Beijing, China}
}

% \vspace{-0.1pt}

\author{Xiaohui Zheng}
\email{zhengxiaohui@ict.ac.cn}
\affiliation{
  \institution{ICT, CAS}
  \country{Beijing, China}
  % \location{Beijing, China}
}

% \vspace{-0.1pt}

\author{Xiaofang Zhao}
\email{zhaoxf@ict.ac.cn}
\affiliation{
  \institution{ICT, CAS}
  \institution{IICT, Suzhou, CAS}
  \institution{UCAS, Nanjing} 
  \institution{Zhongguancun Lab}
  \country{Beijing, China}
  % \location{Beijing, China}
}

\renewcommand{\shortauthors}{Cunchi Lv et al.}

\renewcommand{\shorttitle}{\projectname{}: Enabling GPU Resourcing-on-Demand for Serverless DL Serving via IE}

\begin{abstract}
Serverless computing, with its ease of management, auto-scaling, and cost-effectiveness, is widely adopted by deep learning (DL) applications. DL workloads, especially with large language models, require substantial GPU resources to ensure QoS. However, it is prone to produce GPU fragments (e.g., 15\%-94\%) in serverless DL systems due to the dynamicity of workloads and coarse-grained static GPU allocation mechanisms, gradually eroding the profits offered by serverless elasticity.

% To provide holistic elasticity and enhance serving throughput, we present Dilu, an introspective, 2-dimension-scaling serverless DL system designed for GPU resourcing-on-demand GPU. It removes fragments by collocating resource complementary DL instances and dynamically adjusting GPU provisioning. Dilu seamlessly integrates scaling up/down and scaling in/out, significantly enriching the elasticity of the serverless DL system. Our prototype system tested in a local cluster, boosts inference throughput by up to 3.8$\times$ and training throughput by 2.5$\times$, and reduces GPUs by 30\% in large-scale simulations, outperforming the state-of-the-art baseline.
Different from classical serverless systems that only scale horizontally, we present introspective elasticity (IE), a fine-grained and adaptive two-dimensional co-scaling mechanism to support GPU resourcing-on-demand for serverless DL tasks. Based on this insight, we build \projectname{}, a cross-layer and GPU-based serverless DL system with IE support. First, \projectname{} provides multi-factor profiling for DL tasks with efficient pruning search methods. Second, \projectname{} adheres to the resourcing-complementary principles in scheduling to improve GPU utilization with QoS guarantees. Third, \projectname{} adopts an adaptive 2D co-scaling method to enhance the elasticity of GPU provisioning in real time. Evaluations show that it can dynamically adjust the resourcing of various DL functions with low GPU fragmentation (10\%-46\% GPU defragmentation), high throughput (up to 1.8$\times$ inference and 1.1$\times$ training throughput increment) and QoS guarantees (11\%-71\% violation rate reduction), compared to the SOTA baselines.

% To provide holistic elasticity and enhance serving throughput, we present Dilu, an introspective, 2-dimension-scaling serverless DL system designed for GPU resourcing-on-demand GPU. It removes fragments by collocating resource complementary DL instances and dynamically adjusting GPU provisioning. Dilu seamlessly integrates scaling up/down and scaling in/out, significantly enriching the elasticity of the serverless DL system. Our prototype system tested in a local cluster, boosts inference throughput by up to 3.8$\times$ and training throughput by 2.5$\times$, and reduces GPUs by 30\% in large-scale simulations, outperforming the state-of-the-art baseline.

\end{abstract}





\newcommand{\Projectname}{Dilu}
\newcommand{\projectname}{\Projectname{}}

\settopmatter{printfolios=false}

\maketitle

\section{Introduction}

\begin{figure}
    \centering
    \setlength{\abovecaptionskip}{0.cm}
    \includegraphics[scale=0.4]{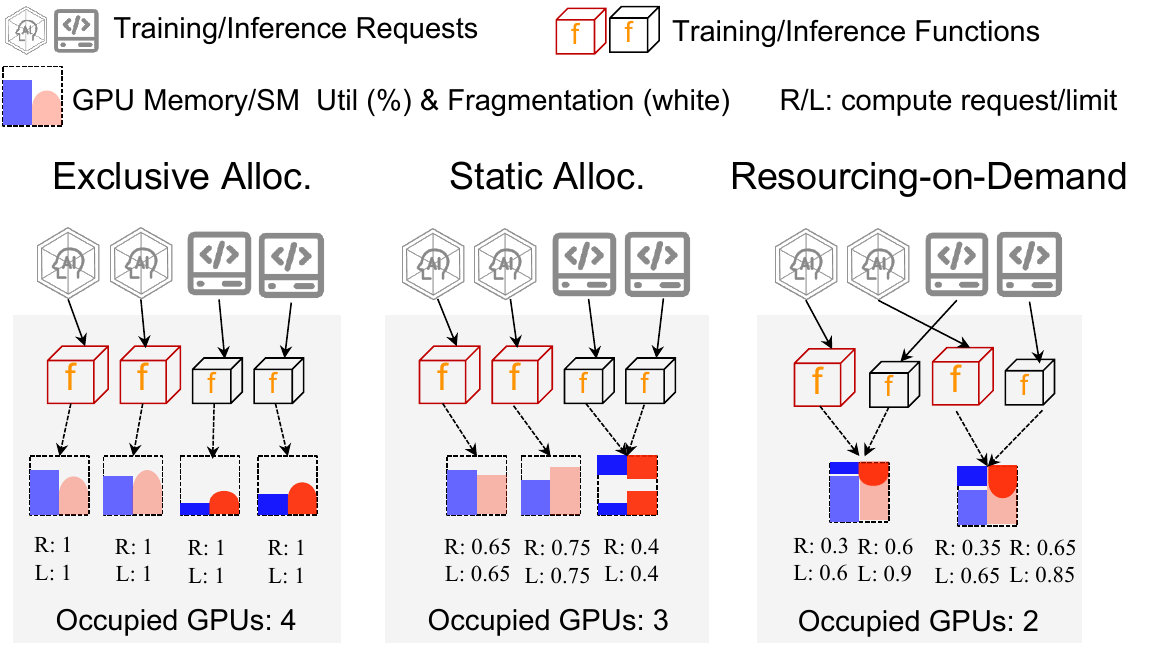}
    \caption{GPU provisioning of serverless DL systems.}
    \label{fig:mechanisms}
    % \vspace{-0.1in}
\end{figure}

Serverless computing has been widely used in DL serving. Many end-to-end DL platforms, like AWS SageMaker \cite{AmazonSageMaker}, Alibaba PAI \cite{PAI} and Microsoft ACI \cite{ACI}, have adopted serverless concepts to provide ease-of-use experience and reduce management efforts of developers. Studies investigate the integration of elastic training \cite{elasticflow,lambdaml,lambdadnn} and inference  \cite{infless,amps,fastg} tasks within serverless, offering numerous benefits such as low resource consumption, automatic deployment, and auto-scaling.
 Propelled by the emergence of Large Language Models (LLMs), GPU-based serverless DL systems become more popular and notable \cite{infless,fu2024serverlessllm,fastg,mlsyssla}.

% Driven by advances and large scale in deep learning (DL), DL serving in the cloud has rapidly evolved. Serverless computing, with its simplified development management, high scalability, and cost-effective pay-as-you-go pricing model, has become the preferred serving platform for DL developers and algorithm enterprises \cite{azure,PAI,AmazonSageMaker}.

% However, As models, especially Large Language Models (LLMs), grow continually in size, traditional CPU computing struggles to ensure service quality, leading to unsatisfactory performance. Consequently, GPUs and other accelerators have become a necessity for DL serving in the new era. However, there is a significant gap between the current serverless DL systems' mechanism of bulky GPU resource provisioning and their inherent advantages in ultra-lightweight elasticity. As shown in Figure~\ref{fig:mechanisms}, most systems  \cite{ACI,elasticflow,fu2024serverlessllm,mlsyssla} provide each function instance with an exclusive GPU, regardless of the extent of fragmentation, resulting in a significant waste of costly GPU resources. To improve resource utilization, some works like  \cite{infless,fastg}, leverage slightly fine-grained GPU provisioning via MPS \cite{NVIDIAMPS}, allowing multiple \textit{homogenous} instances to share a single GPU. However, these static allocation mechanisms fail to match the dynamic nature of DL workloads, where the required GPU resource sizes constantly fluctuate. 
However, GPU fragmentation tends to occur in serverless DL systems due to various factors, such as the dynamicity of DL task workloads, static GPU allocation and keep-alive strategies to balance the overheads of cold starts. Further, it leads to several issues, such as low resource utilization, and high cost, which significantly undermine elasticity and cost-efficiency (i.e., pay-as-you-go) brought by serverless computing. Specifically, as shown in Figure~\ref{fig:mechanisms}, most systems  \cite{elasticflow,fu2024serverlessllm,mlsyssla} adopt the exclusive GPU allocation method (left), resulting in a significant waste of GPU resources. Works like \cite{infless,fastg,dhakal2020gslice}, leverage slightly fine-grained GPU provisioning via MPS \cite{NVIDIAMPS}, allowing several instances to share a single GPU 
with a fixed resource quota (medium), which lacks efficient dynamic resourcing adjustment. We argue that resourcing-on-demand GPU provisioning is the ideal status to improve GPU utilization in serverless computing (right), maximizing the potential benefits of serverless elasticity.

% We observe that both DL training and inference workloads have dynamic characteristics like Figure~\ref{fig:timeline}, leading to each instance not consistently fully utilizing the allocated GPU resources, thus resulting in uneven GPU idling. This dynamicity poses an enormous challenge to the elastic resource provisioning typical of serverless architectures. Thus, we argue that an ideal serverless DL system should provide holistic flexibility, rather than merely focusing on horizontal auto-scaling.

Constrained by the aforementioned exclusive or static GPU provisioning, existing serverless DL systems \cite{fastg,infless,elasticflow} merely scale horizontally (i.e., scaling in/out at the inter-instance level) reactively in response to workload changes. In contrast, we present introspective elasticity (IE), which means fine-grained and dynamic GPU provisioning for serverless DL functions, supporting vertical scaling (scale up/down GPU compute cores at the intra-instance level) and horizontal scaling in a coordinated manner to minimize GPU fragmentation. However, it is non-trivial to design such a cross-layer system. It faces several challenges, including accurately profiling basic resource requirements of DL tasks to identify fragments, scheduling with consideration of multiple factors (e.g., Quality of Service/QoS, heterogeneous workloads), and performance interference caused by disordered resource consumption between instances in real time.
% \textit{It is built with 2-dimensional scaling to realize resource scaling from fine-grained to corase-grained with least fragments and real-time resource provisioning. It raises three challenges to acquire such elasticity, including the profiling of resourcing requirements, scheduling with collocations to reach multiple goals, and scaling with low resource contention.} 

% To enhance utilization and aggregate throughput of the cloud cluster, we develop \projectname{}, an introspective, multi-dimension elastic serverless DL system designed to achieve GPU resourcing on demand. First, to identify the temporal and spatial fragmentation of heterogeneous DL instances, we introduce efficient pruning strategies (i.e., binary search for training and a hybrid growth strategy for inference). Further, our introspective resourcing complementary scheduling minimizes the number of occupied GPUs while ensuring the Quality of Service (QoS) for each DL task. Most importantly, our proposed Adaptive 2D Co-Scaling strategy integrates intra-instance-level scaling up/down with inter-instance-level scaling in/out, providing holistic elasticity for the serverless DL system. This mechanism smoothly transitions during bursty workloads and significantly reduces the frequency of cold starts.

In this paper, we build \projectname{}, a serverless DL system designed to achieve GPU resourcing-on-demand with introspective elasticity.
First, to determine multiple resourcing requirements to facilitate identifying GPU fragments, especially GPU compute resources, we introduce efficient pruning search strategies for heterogeneous DL tasks. Second, we design a resourcing-complementary scheduling policy to minimize GPU consumption while ensuring QoS of collocated tasks. Most importantly, we present an adaptive two-dimensional co-scaling strategy, combining fast scaling-up/down with lazy scaling-out/in. It can dynamically adjust GPU provisioning and smoothly transition between vertical scaling and horizontal scaling, promoting DL serving performance. The evaluation shows that 
\projectname{} successfully delivers GPU resourcing-on-demand for serverless DL tasks, reducing fragmentation by 10-46\% and boosting inference and training throughputs by $1.8\times$ and $1.1\times$ compared to the SOTA baseline. It also guarantees QoS by reducing 11-71\% violation rate.
% \projectname{} can effectively provide GPU resourcing-on-demand for serverless DL functions with 10\%-46\% lower GPU fragmentation, 1.8$\times$ and 1.1$\times$ higher throughput of inference and training respectively than INFless+, and QoS guarantees with 11\%-71\% less violation rate. 

In summary, we make the following major contributions:
\begin{itemize}[leftmargin=*,itemsep=0pt,parsep=0pt,topsep=1pt,partopsep=1pt]
    % \item We observe that existing serverless DL systems sustain underutilized GPU usage and propose to build an introspective serverless DL system with GPU resourcing-on-demand elasticity.
    % \item We study GPU fragmentation of current serverless DL systems and propose to build an GPU resourcing-on-demand system with introspective elasticity.
    
    \item We present an efficient binary-search-based GPU resource profiling method for training and a Hybrid Growth Search Strategy for inference tasks, where the latter speeds searching efficiency up to 3.3$\times$ compared to the SOTA method.
    % \item We propose pruning-search based multi-factor profiling methods. It improves the efficiency by up to 3.3$\times$ compared to the SOTA baseline.

    \item We introduce a resourcing-complementary scheduling method to defragment GPU while considering QoS, significantly improving GPU utilization and increasing function deployment density.
    % improve function deployment density xxxxxxxx.

    % \item We co-design scaling up/down and scaling in/out to achieve holistic elasticity. \projectname{} reduces the cold start rate by 91\% at most, while maintaining the lowest QoS violation rate.
    \item We co-design fast scaling-up/down and lazy scaling-in/out to achieve introspective elasticity in a cross-layer manner. It reduces the cold start rate of inference functions by 91\% at most, while maintaining the lowest serving violation rate.
 
    \item We develop a prototype system of \projectname{} on Kubernetes and Docker, which is publicly available\footnote{\href{https://github.com/sigserverless/Dilu}{https://github.com/sigserverless/Dilu}.}. The cluster evaluations show that \projectname{} boosts inference and training throughputs by 1.8$\times$ and 1.1$\times$ compared to INFless.
    
\end{itemize}

\section{Background and Motivation}
\label{Sec:Background-and-Motivation}

\begin{figure*}[t]
    \centering
    \setlength{\abovecaptionskip}{0.cm}
    % \captionsetup[subfigure]{skip=0pt} 
    % \captionsetup[subfigure]{skip=-3pt} 
    \subfigure[Temporal workload characteristics ]{
        % \captionsetup[subfigure]{skip=-2pt}
        \includegraphics[scale=0.415]{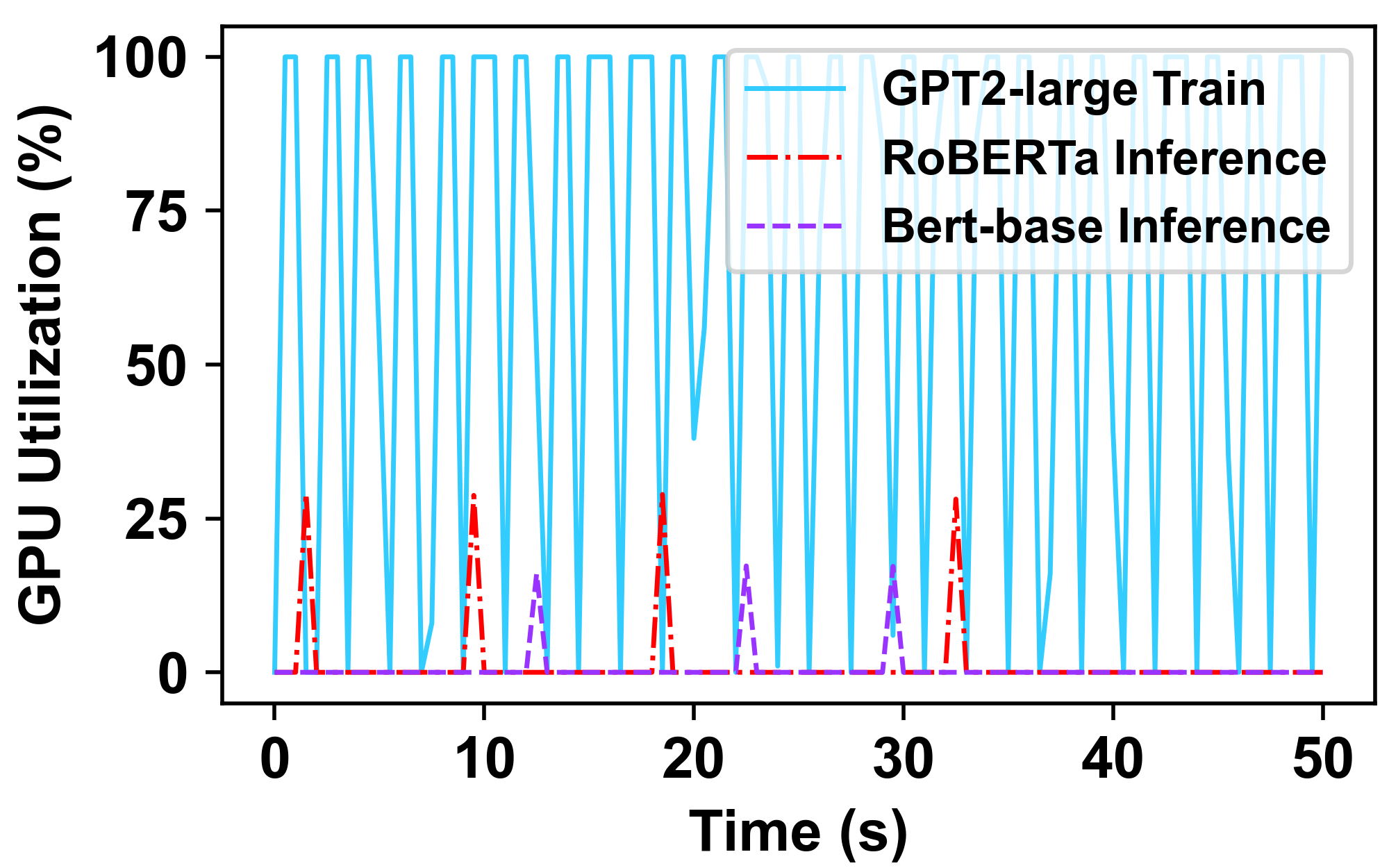}
        \label{fig:timeline}
    }
    \hspace{-0.01\linewidth} % Adjust spacing between subfigures
    \subfigure[Spatial occupying characteristics]{
        \includegraphics[scale=0.415]{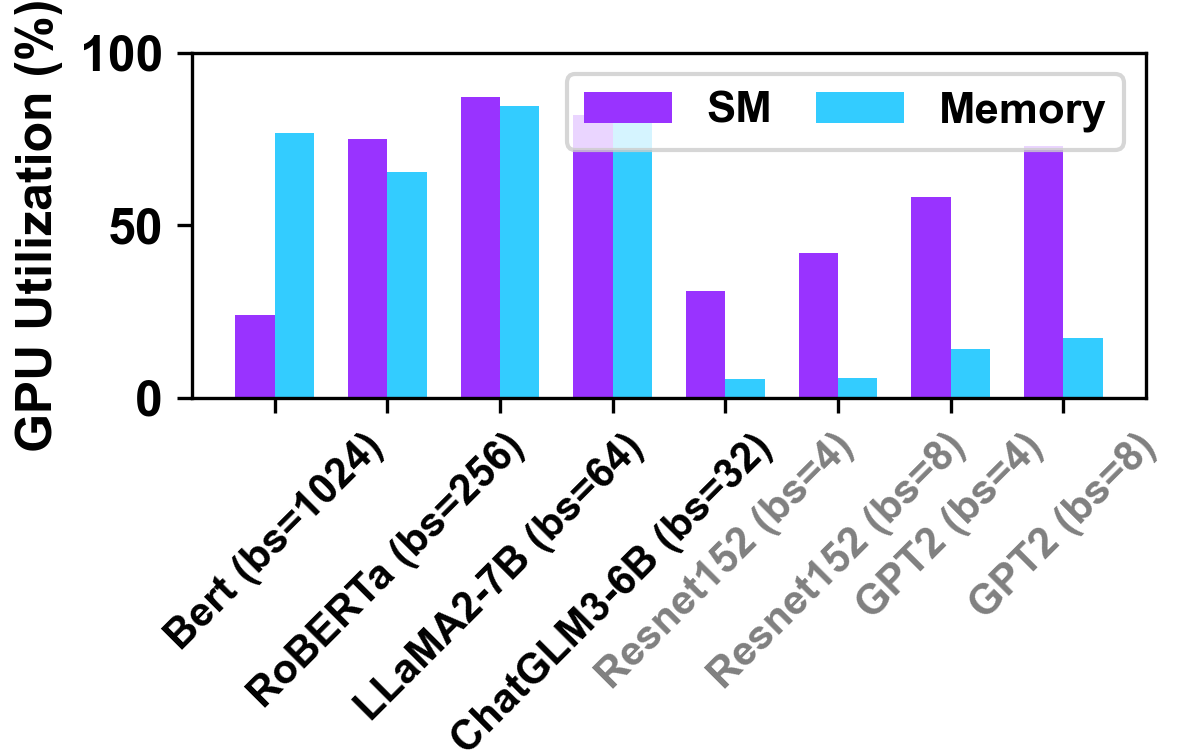}
        \label{fig:spatial-feas}
    }
    \hspace{-0.01\linewidth}
    \subfigure[Inference latency]{
        \includegraphics[scale=0.415]{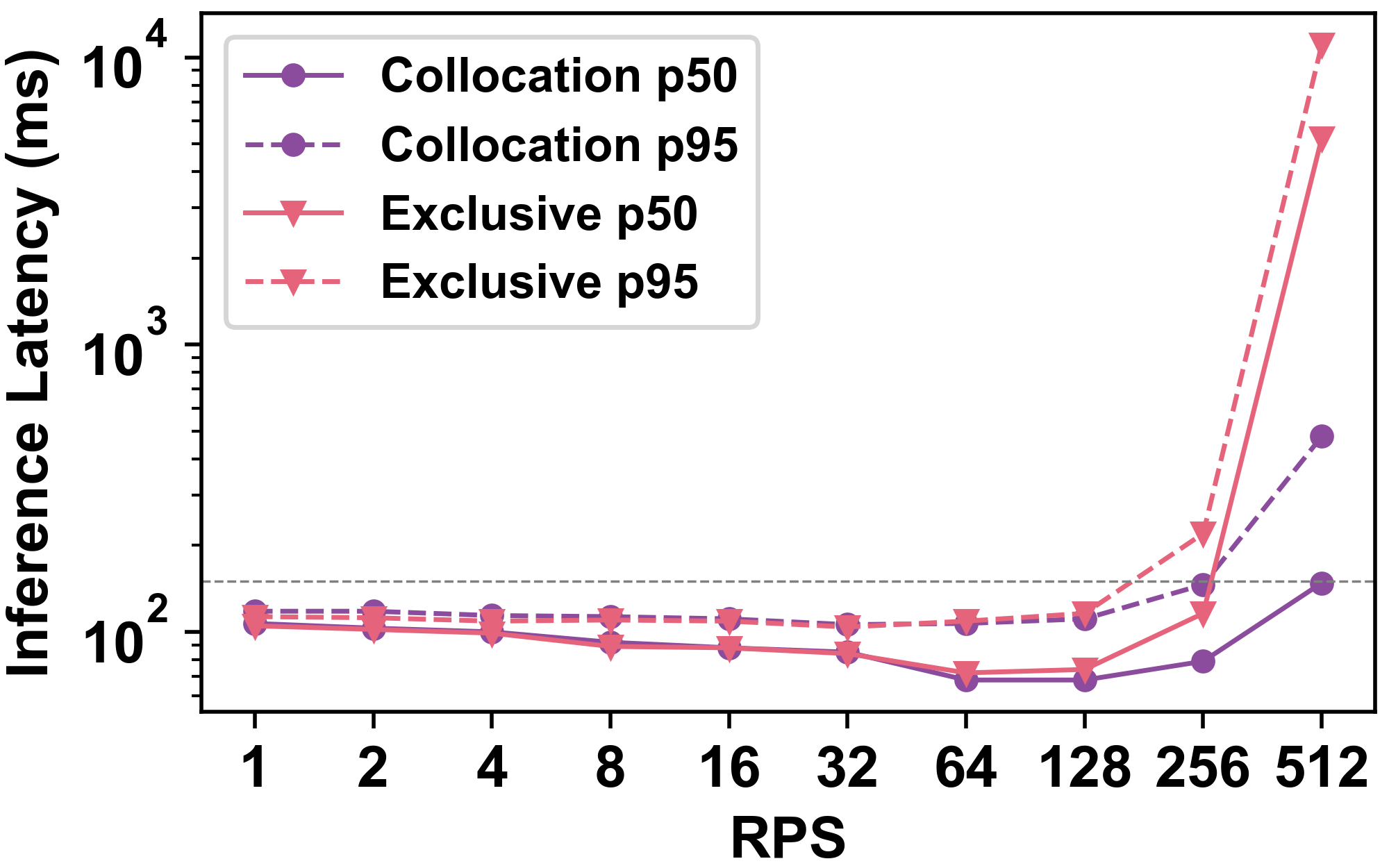}
        \label{fig:toy-inf}
    }
    \hspace{-0.01\linewidth}
    \subfigure[Training throughput]{
        \includegraphics[scale=0.415]{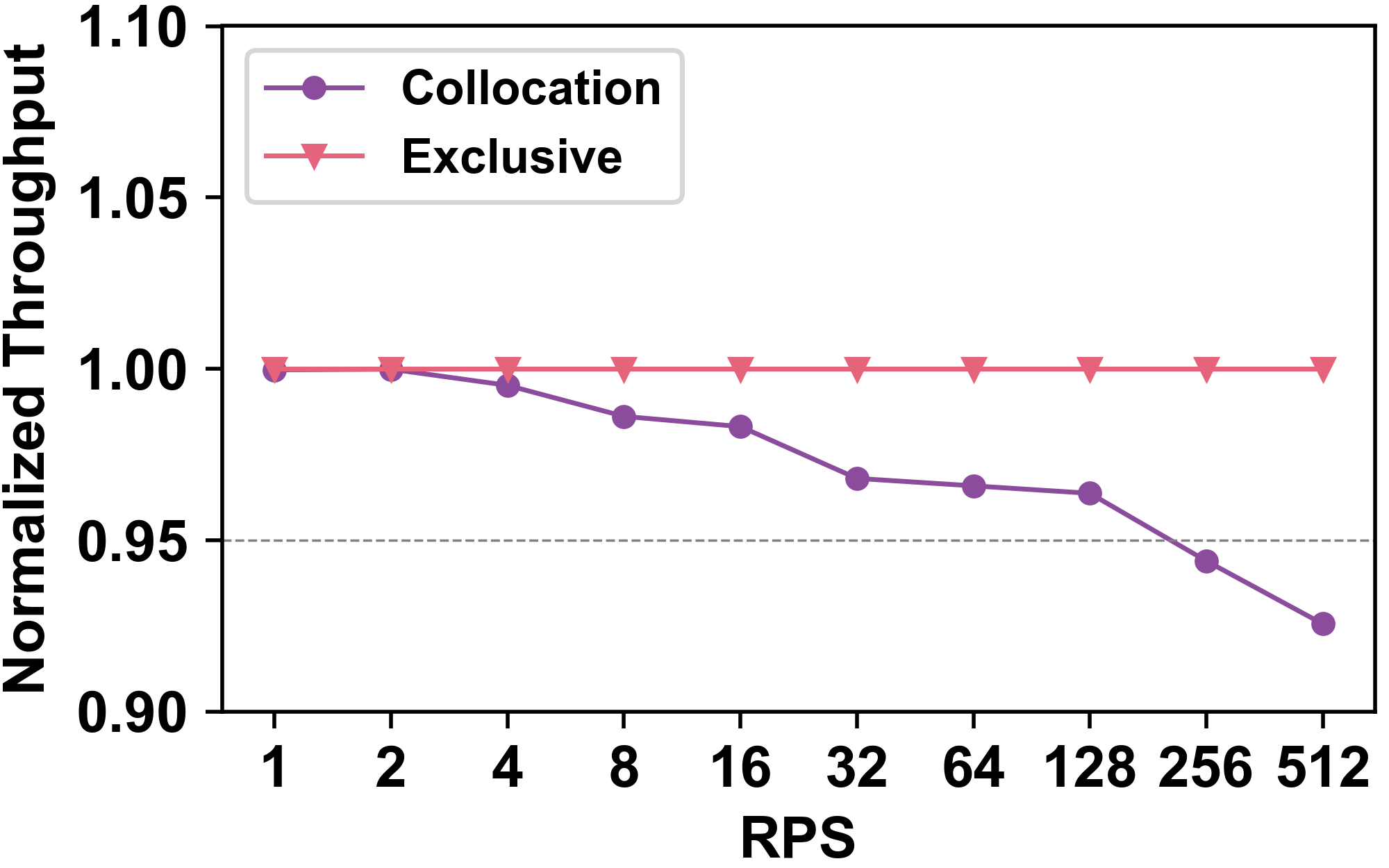}
        \label{fig:toy-train}
    }
    % \captionsetup{skip=0.2pt}
    \caption{Observations on serverless DL serving. (a)(b): GPU fragmentation in temporal and spatial dimensions. The black-color models represent training and the gray is for inference models. (c)(d): The RoBERTa-large inference latency and Bert-base training throughput comparisons under the co-scaling mechanism using 3 GPUs, relative to the Exclusive mode using 4 GPUs. }
    \label{fig:background-characteristics}
    % \vspace{-0.1in}
\end{figure*}

% In this section, we first introduce briefly training and inference tasks in terms of using GPU, and then indicate limitations of existing solutions for GPU sharing. Finally, based on the characteristics of heterogenous DL tasks, we point out our motivation.

\subsection{Serverless DL Serving}\label{subsec:serverless}
Considering manifold drawbacks (e.g., high resource consumption, complicated server operations) of server-centric DL serving, cloud providers and studies promote the serverless DL serving patterns and deliver explicit progress. Nowadays, propelled by LLMs, serverless DL serving becomes more promising to provide elastic GPU provisioning. % In this section, we summarize the status and tendency of both serverless training and inference. End-to-end DL platforms, like AWS SageMaker\cite{AmazonSageMaker}, Alibaba PAI\cite{PAI} and Microsoft ACI\cite{ACI}, have widely adopted serverless concepts to provide ease-to-use experience, reducing management efforts of developers.

%\textbf{Serverless training}. 
%DL training involves iteratively updating the parameters of a deep neural network (DNN) by progressively narrowing the discrepancies between predictions and ground-truths, typically involving forward and backward propagation steps. Distributed training seeks to accelerate the process by distributing the workload across multiple GPUs or nodes. In the industry, the introduction of Serverless Platforms (e.g., AWS SageMaker\cite{}, Alibaba PAI\cite{}, and Microsoft ACI\cite{}) has provided developers with highly elastic, cost-effective, and user-friendly DL computing services. Concurrently, numerous academic efforts have propelled advancements in this field. In terms of elastic scaling, ElasticFlow\cite{} leverages residual cluster GPU resources and model scaling curves to periodically adjust the number of DDP workers, while lamdaDNN\cite{} dynamically adjusts worker numbers based on model convergence to minimize training iterations. For deployment simplification, some works ( e.g., LambdaML\cite{}, Siren\cite{}, and Cirrus\cite{}) focus on one-stop deployment, whereas FuncPipe\cite{} and Hydrozoa\cite{} concentrate on automating model partitioning for hybrid parallel training. In terms of cost-reduction, Dorylus\cite{} utilizes the elastic scaling capabilities of Serverless functions to handle extensive GNN computations.

\textbf{Serverless Training}.
% Cloud providers benefits from the serverless paradigm to realize ready-to-use online training services, automatic deployment of general or customized training workflows, and elastic scaling for basic or adaptive training. Multiple parallelism strategies are usually adopted to manage forward and backward propagation of training processes across multiple GPUs or nodes to achieve acceleration. The employing and deploying can be lengthy, and the optimization, like elastic training, can be challenging.  
Providers or developers benefit from the serverless paradigm to build ready-to-use online training services, automatic deployment of training workflows (e.g., LambdaML \cite{lambdaml}, Siren \cite{siren},  Cirrus \cite{cirrus}, FuncPipe \cite{funcpipe}, Hydrozoa \cite{hydrozoa}), and elastic scaling for training workers (e.g., ElasticFlow \cite{elasticflow}, $\lambda$DNN \cite{lambdadnn} and Dorylus \cite{dorylus}). Specifically, FuncPipe \cite{funcpipe} and Hydrozoa \cite{hydrozoa} automate model partitioning for dynamic hybrid parallel training, and ElasticFlow \cite{elasticflow} adaptively adjust serverless DP workers according to residual GPU resources.

\textbf{Serverless Inference}.
% Unlike training, inferences are mainly online, and the workloads usually fluctuate, which is naturally suitable with the classic serverless paradigm. Cloud providers implement serverless inference services to help developers rapidly build elastic inference services and deliver online API invocations in various scenarios. Since the serving process is more time- and cost-sensitive, studies explore providing stringent Service Level Objectives (SLO, e.g., latency within 100ms) guarantees, reducing redundant resourcing costs and cold-start overheads.
% Unlike training, inferences are mainly online with fluctuated workloads, naturally suitable with the  serverless paradigm. It can help developers rapidly build elastic inference services with API invocations. Since the inference is more time- and cost-sensitive, studies pay attention on guaranteeing Service Level Objectives (SLO, e.g., latency within 100ms), reducing redundant resourcing costs and cold-start overheads.
Unlike training, inference is mainly online with fluctuated workloads, naturally suitable with the serverless paradigm. Since inference is more time-sensitive, studies \cite{gillis,vldb_inf,naranjo2020accelerated, amps,mark2019,tetris} pay attention to guaranteeing Service Level Objectives (SLOs, e.g., inference latency within 100ms). Additionally, studies like INFless \cite{infless} and BATCH \cite{batch} introduce batching execution to serverless to improve GPU utilization. Others \cite{fu2024serverlessllm,infless,serverlessinthewild} leverage layered caches and keep-alive strategies to reduce cold start overhead.
% Studies\cite{gillis,vldb_inf,naranjo2020accelerated, amps} leverage or improve serverless elasticity to handle bursty workloads with SLO guarantees. For reducing costs, studies\cite{mark2019,tetris} optimize the usage of instance resources such as CPU and host memory across multiple inference instances to save resource allocations. Additionally, studies like InfLess\cite{infless} and BATCH\cite{batch} let inference functions handle batching requests simultaneously to improve GPU utilization. To reduce cold starts, studies like ServerlessLLM\cite{fu2024serverlessllm} and INFless\cite{infless}, leverage layered cache of DL models and keep-alive strategies respectively, to cut down the cold starts of inference instances.

% Studies\cite{gillis,vldb_inf,naranjo2020accelerated, amps} handle bursty workloads with SLO guarantees by leveraging serverless elasticity. For reducing costs, studies\cite{mark2019,tetris} optimize the usage of instance resources such as CPU and host memory across multiple inference instances to save resource allocations. Additionally, studies like InfLess\cite{infless} and BATCH\cite{batch} introduce batching techniques, grouping multiple inference requests and executing simultaneously, to increase resource utilization. To reduce cold starts, studies like ServerlessLLM\cite{fu2024serverlessllm} and INFless\cite{infless}, leverage layered cache of model parameters and prediction-based keep-alive strategies respectively, to cut down or remove cold start frequency of inference instances, improving service efficiency with lower cost.

\textbf{Trends}.\label{subsec:trends}
% Serverless has been widely used in DL serving. However, most recent studies primarily support serverless DL serving with CPUs, like\cite{tetris,batch,vldb_inf,funcpipe,amps,mark2019,cirrus,siren,lambdadnn}. They are not well-suited to complex DL tasks, especially for LLM-based workloads. Cloud providers\cite{AmazonSageMaker,ACI,PAI} have integrated the platforms with GPU support. Unlike CPU-based serverless DL serving with function-based small services, the computing resource intensity of current DL workloads increases. It forms larger serverless functions, relying on GB-scale model copy and loading in cold-start, and more and longer resource usage of GPUs. The prominent changes ask for fine-grained GPU resource management for maintaining and exploring. We will further discuss it in \textbf{section} \ref{subsec:fragments}.
% Serverless has been widely used in DL serving. However, most studies explore with CPUs, like\cite{tetris,batch,vldb_inf,funcpipe,amps,mark2019,cirrus,siren,lambdadnn}, not well-suited to complex DL tasks, especially the LLM-based ones.
The serverless DL functions are more compute- or memory-bound due to the larger model sizes. Though studies \cite{fu2024serverlessllm,infless,fastg,elasticflow,mlsyssla} devote to building GPU-based serverless DL systems, they suffer from coarse-grained GPU resourcing techniques, which hinders their ability to deliver highly efficient, elastic, and cost-effective DL services. % We will further discuss it in \textbf{section} \ref{subsec:fragments}.

% Currently, most Serverless platforms (e.g., ) primarily support CPUs, which are not well-suited for complex DL workloads. To accelerate training and inference, some works(e.g., ) have integrated GPU resources into Serverless, providing GPUs to DL tasks in an exclusive or fixed quota manner. This approach often results in severe overprovisioning, which we will discuss in detail in the next section. LLM.

% \cite{mark2019,gillis,infless,nacl_inf,naranjo2020accelerated,tetris,amps, batch,dgsf,mlsyssla,vldb_inf,wu2022serverless} make full use of the elasticity of serverless to cope with unpredictable and burst inference workloads. \cite{amps, gillis} supports model-parallelism inference. \cite{mark2019, tetris, batch,vldb_inf} mainly focuses on the optimization of CPU and host memory costs. Among these, \cite{dgsf,infless,naranjo2020accelerated,faaswap,fastg,reducinginfcold,mlsyssla} introduce GPUs to accelerate serving. Works like \cite{choi2022serving, infless,batch, vldb_inf} improve execution efficiency by batching. \cite{fastg,infless} leverage MPS to share GPU, improving GPU utilization. Compared with \projectname{}, the studies focus on either training or inference only and are unable to remove the notable GPU underutilization in fine-grained ways.

\subsection{GPU Resourcing}\label{subsec:GPU}
% GPUs are key devices for DL tasks, supporting the compute, memory and communication requirements. In this section, we summarize the usage of GPU devices, GPU sharing, and explain the GPU resourcing-on-demand purposes.

%\textbf{GPU Devices}.
%GPUs are increasingly prominent in parallel computing and have become highly sought-after and scarce resources due to the surge in deep learning, particularly in Large Language Models (LLMs). This context demands a maximization of GPU utilization. The primary performance metrics for GPUs focus on GPU memory and Streaming Multiprocessors (SMs), each equipped with numerous Tensor Cores, CUDA Cores, and Registers. In DL, the memory size determines the model capacity a GPU can accommodate, and without memory swapping, it typically does not impact the training or inference performance.  SMs are pivotal in computing processes. Algorithm engineers program DL models using high-level frameworks like PyTorch\cite{} or TensorFlow\cite{}, subsequently compiling them into CUDA kernels for execution on SMs. The initiation and dispatch of these CUDA kernels are managed by CPUs. However, due to the proprietary nature of drivers (e.g., NVIDIA Driver), it is impractical to directly control the execution order of kernels across applications on GPU SMs (although kernels within a single application execute sequentially) or to allocate SM resources dynamically. This limitation also adds complexity to the dynamic and proportional allocation of computational resources across tasks.

\textbf{GPU Device.}
% With the emergence of LLMs, GPUs are increasingly prominent in DL computing. GPUs (e.g., Nvidia\cite{NVIDIA}) mainly consist of several Streaming Multiprocessors (SMs), each equipped with numerous Tensor Cores, CUDA Cores, Registers, and a bunch of memory. CUDA kernels, containing groups of threads, are issued to SMs in parallel, and access data from memory. In terms of DL tasks, high-level DL programs (e.g., with PyTorch\cite{PyTorch} or TensorFlow\cite{TensorFlow} frameworks) will be transformed into CUDA kernels and dispatched to run on SMs, the DL models will be loaded into GPU memories for the kernels to access in runtime. However, the computation characteristics of DL tasks change in types or stages. For example, forward and backward propagation steps in training are compute-bound, while the gradient synchronization stage is memory-bound(TODO-3 reference). The encoding stage in LLM inferences is compute-bound(i.e., cublasGemmEx) while the token generation is memory-bound(i.e., max\_pooling,layer\_norm\_fwd, reference). It indicates that the resource demand trends of DL tasks in runtime are quite different. However, the low-level resources in GPU devices are not straightforward to schedule or control. The practical effect relies on CUDA library implementation, CUDA GPU drivers, and GPU architecture. Thus, it is hard to allocate GPU resources on demand to fit with the real-time requirements of DL tasks.
With the LLM emergence, GPUs are increasingly prominent for DL. GPUs consist of several Streaming Multiprocessors (SMs), each equipped with numerous Tensor Cores, CUDA Cores and scarce memory. High-level DL programs (e.g., based on PyTorch \cite{PyTorch}, TensorFlow \cite{TensorFlow}) are first transformed into CUDA kernels and dispatched to run on SMs. However, DL tasks vary significantly in types and sizes of kernels. 
For example, forward and backward propagations in training are compute-bound, while gradient synchronization is usually memory-bound caused by communication. The prefilling in LLM inference is compute-bound while decoding is memory-bound \cite{zhong2024distserve}. Thus, it may directly lead to the underutilization of specific GPU resources with improper task assignment at the high level.

% Within a GPU, memory, computation (and bandwidth) indicate its capabilities. More GPU memory (i.e., 40GB in NVIDIA-A100) means it can provide larger space for intermediate data, thus reducing transferring with CPU main memory. Computation resources, e.g. SMs(Streaming Multiprocessors) or CUDA Cores (i.e., 6912 in A100 ), determine the peak floating-point operations per second(FLOPS) achievable, impacting its compute capacity and efficiency.

\textbf{GPU Allocation in Cloud.}\label{subsec:GPU-Sharing} 
% It is a main concern to improve GPU utilization in recent studies. A common method is exclusively allocating GPUs to a single DL task, commonly seen in many serverless DL systems\cite{elasticflow,hydrozoa,fu2024serverlessllm,mlsyssla,faaswap}. Studies also propose GPU sharings to multiplex GPUs in temporal or spatial manners, including hardware-based GPU partition (e.g.,\cite{fastg, dhakal2020gslice,choi2022serving},[????????duibudui] MPS\cite{NVIDIAMPS, infless}), virtual GPU\cite{naranjo2020accelerated}, CUDA interception TGS\cite{tgs,gaiagpu,fastg} and rCUDA\cite{faaswap}. The hardware-based, MPS-based and vGPU methods are usually spatial-sharing, providing fixed partitions among tasks with static quotas. CUDA and rCUDA interception methods can be dynamic temporal sharing, allowing to launch different amounts of CUDA kernels in different stages of same task with different GPU quotas so that idle GPUs. Also, the static and dynamic methods can be combined. The GPU-sharing methods, especially the CUDA interception, are quite instructive. However, they are not carefully co-designed with serverless DL systems. TGS[refxxxxxxxxxx] only considers \textit{training-training} sharing cases. Current studies mainly adopt the exclusive or MPS methods(e.g., in INFless\cite{infless}). The GPU sharing methods for serverless DL servings can be further improved, especially with LLMs, as shown in figure 7, 8, 9 xxxxxxxxxx.
A common method is exclusively allocating the whole GPU to each DL instance, commonly seen in many serverless DL systems \cite{elasticflow,hydrozoa,fu2024serverlessllm,mlsyssla,faaswap}. To further improve GPU utilization, studies leverage GPU-sharing methods to multiplex GPUs, including MPS-based spatial partition used by \cite{fastg, dhakal2020gslice,infless}, virtual GPU \cite{naranjo2020accelerated}, temporal methods used by \cite{tgs,gaiagpu,fastg} and rCUDA \cite{faaswap}. Studies have also explored spatio-temporal sharing methods \cite{choi2022serving, fastg, han2024inss}, all of which are based on MPS. These methods require frequent adjustments of partition sizes at the process level to accommodate the highly fluctuating workloads, leading to significant time overhead. Moreover, due to the static allocation enforced by MPS, they are unable to exceed the resource limits of a single instance to handle burst workloads instantaneously. Thus, a non-negligible gap exists to support fine-grained GPU provisioning on demand in current serverless DL systems.

\subsection{Fragmented GPU Resourcing in Serverless} \label{subsec:fragments}

With current monotonous elasticity, existing serverless DL systems are prone to produce GPU fragments.  % Figure~\ref{fig:spatial-feas} illustrates the SMs and memory utilization of training and inference functions, demonstrating underutilization in the spatial dimension.(????? xxxxx) 
We make the following observations of the fragment sources:

% which further leads to increasing computational costs and a deployment-density reduction\cite{} of GPU functions. Specifically, we make the following observations.
% The size of DL serverless functions is large. expements and results analysis. Reason analysis.  In CPU serverless computing, a function is small and fine-grained using CPU resources, it helps to reduce the fragements. While in GPU environments, a function can be large and consuming GPU intensively. Thus, the fragments is difficult to fill with a task in GPU level. and the fragements can be accumulated at cluster-level. The limitations are mainly from 2 aspects.

% \textbf{Observation-1: Static GPU over-provisioning mismatches the dynamic workloads, yielding temporal and spatial fragments}.

\textbf{Observation-1: GPU overprovisioning.}\label{subsec:gpu-overprovisioning} 
% Serverless platforms easily provide GPUs statically in an over-provisioning manner, mismatching dynamic workloads and losing the opportunity to reuse fragments. Specifically, to ensure the training Job Completion Time(JCT) and the inference SLOs, Serverless platforms allocate the whole GPU or a fixed high rate for DL instances shown in Table~\ref{table:gpuprovision}. For example, INFless scheduler allocates a constant 30\% SM rate for the Bert-base inference instance, although no requests arrive during the keep-alive period (35-50s timespan as shown in Figure~\ref{fig:timeline}), these 30\% SMs are not used by other instances.  Currently, to our knowledge, no method supports unequal requests/limits settings to accommodate dynamic DL workloads. 
% Serverless platforms usually allocate GPU resources statically to DL functions, which are easily over-provisioned with a empirical or profiled quota to ensure the training Job Completion Time(JCT) or the inference SLOs. In resourcing view, it represents an euqal setting of request and limit quotas, as shown in Table~\ref{table:gpuprovision}. For example, INFless allocates a constant 30\% SM rate for the Bert-base inference function. If the CUDA kernels are low compute-bound and the workload is at low-level, the 30\% SMs can be idle as the instances are alive. To our knowledge, no systems support unequal requests\&limits settings for serverless DL serving yet.
With static GPU allocation for DL functions, each instance is easy to be over-provisioned since the resource quotas are often set empirically high or assigned in a coarse-grained manner to ensure training job completion times or meet inference SLOs. As Table~\ref{table:gpuprovision} shows, taking the basic resource definition with \textit{<request, limit>} in Kubernetes as an example, both Exclusive- and MPS-based allocation mechanisms can be regarded as an equal setting of \textit{request} and \textit{limit} quotas. Specifically, as shown in Figure~\ref{fig:timeline}, INFless allocates a constant 30\% SM rate to handle RoBERTa inference with a batch size of 4 at maximum, while this SM quota cannot be reallocated to other instances dynamically, though the workload is low. Figure~\ref{fig:spatial-feas} shows the average GPU utilization in terms of SM and memory, which is much lower than the actual allocated GPU resource quotas, especially under the exclusive allocation \cite{elasticflow,hydrozoa,mlsyssla}.

\textbf{Observation-2: GPU idling of serverless serving.}\label{subsec:communication-idling}
In distributed training, each worker needs to communicate to synchronize gradients. Since this process does not consume GPU compute resources, it results in significant GPU idling time. As shown in Figure~\ref{fig:timeline}, the GPU idling time exceeds 40\%, in which a 4-worker GPT2-large training task utilizes PyTorch.DDP \cite{DDP} and NCCL \cite{nccl}. We also fine-tune LLaMA2-7B with pipeline parallelism via DeepSpeed \cite{DeepSpeed} and each worker yields nearly 20\% GPU idling SM as shown in Figure~\ref{fig:spatial-feas}. As for distributed inference, works like \cite{fu2024serverlessllm, gillis, amps, funcpipe} introduce model parallelism to serverless, which undoubtedly leads to substantial bubble time due to pipeline execution characteristics.

% Studies[xxxxxxxxx] explore to collocate tasks with them, while easily causing the low priority tasks to be heavily starving, as shown in figure xxxxxx \& refTGS. % It needs a resonable requisition instead of the rough preemption.

\textbf{Observation-3: Keep-alive strategy for serverless inference tasks.}\label{subsec:keep-alive-idling}
Keep-alive is adopted in serverless DL systems \cite{serverlessinthewild,fastg,infless,tetris} to balance cold start overheads, and also becomes an important source to exacerbate fragmentation. % As the keep-alive functions are free, the resources turn to be fragments.
As shown in Figure~\ref{fig:timeline}, we simulated the workload traces according to Faaswap\cite{faaswap}, which indicates that less than 85\% of functions are invoked per minute on average in \textit{Alibaba}. Two keep-alive inference function instances only handle 3-4 requests within a nearly-50s lifecycle.  It means the keep-alive strategy brings over 95\% of resource waste in the time dimension. 

\begin{table}
    \caption{Comparison of GPU provisioning in serverless. \projectname{} allows the definition of unequal \textit{request} and \textit{limit} quotas during profiling and dynamically adjusts resource provisioning based on real-time demand.}
    \small
    \centering
    \begin{tabular}{cccc}
        \toprule
        \textbf{Mechanism} & \textbf{Req/Lim} & 
        \textbf{Allocation} &
        \textbf{Resourcing} \\
        \midrule
        Exclusive & equal,1 & - & static \\
        MPS & equal,<1 & profiling-based  & static \\
        Dilu &  unequal,<1 & profiling-based & On-Demand \\
        \bottomrule
    \end{tabular}

    % \vspace{-0.2in}
    \label{table:gpuprovision}
\end{table}

\textbf{Implications.} \label{subsec:Implications}
The GPU fragmentation and allocation limitation pull down the elasticity and deployment density \cite{du2022serverless} of serverless DL functions, increasing both user and provider costs. We argue that a proper GPU resourcing-on-demand mechanism is imperative and essential for current serverless DL serving. % We argue that GPU resourcing-on-demand for Serverless DL serving can primarily and uniformly reduce the various fragmentation sources.

\subsection{Motivation and Challenges}

\textbf{Insight: Introspective Elasticity.} 
Introspective Elasticity (IE) is a specialized mechanism for GPU resourcing-on-demand. It refers to a holistic and novel GPU provisioning paradigm tailored for serverless DL functions, which provides fine-grained, continuous and adaptive GPU resources. Unlike current horizontal-only elasticity in serverless \cite{infless,fastg}, which merely provides discrete GPU provisioning and focuses solely on eliminating external resource fragments, IE expands it by dynamically multiplexing internal GPU fragments of instances according to real-time kernel-level workloads, to maximize GPU utilization.
% To enable genuine resourcing-on-demand, a novel and cross-layer approach is required, unleashing full elasticity in GPU consumption. 
% \textcolor{red}{Our insight for IE stems from the limitations of existing serverless DL systems, which struggle with either static and inflexible scheduling-level mechanisms (e.g., MPS or exclusive) or blunt runtime-level provisioning (e.g., keep-alive SMs failing to reuse by other instances instantaneously). }

% \textcolor{red}{needs focus on to needs. Specifically, a system with IE support should first identify static GPU fragments and eliminate them by collocating instances at scheduling. During runtime, dynamic fragments generated by varying workloads are reused among collocated instances. When bursty workloads occur, the vertical scaling provided by collocation coordinates with traditional horizontal scaling to handle the load.}

% \textcolor{red}{Specifically, a system with IE support needs to identify static GPU fragments, which can be eliminated by collocating resource-complementary instances at scheduling. It also needs to focus on how to reuse dynamic fragments yielded by varying workloads. More importantly, the coordinated two abilities above can work together to handle more bursty workloads.}

IE requires the system to identify static GPU fragments, which provides the opportunity to eliminate them with the collocation method. It also emphasizes the need to efficiently reuse dynamic fragments generated by varying workloads. More importantly, these two capabilities ought to be coordinated to effectively manage bursty workloads.

\textbf{Preliminary Verification.} We verify this idea through a toy experiment using Exclusive and Collocation setups with serverless functions: the Exclusive setup involves 4 GPUs, 3 GPUs for training and 1 GPU for inference. In contrast, the Collocation mode occupies only 3 GPUs, each collocating a training worker and an inference worker (vertical scaling). For the collocation case, requests are loaded balance to 3 inference workers (horizontal scaling). The results in subfigures~\ref{fig:toy-inf} and ~\ref{fig:toy-train} show that on top of saving 25\% of GPU resources, the co-scaling mechanism effectively improves inference throughput by 46\% with QoS guarantees while merely decreasing throughout of the collocated training task by 5.2\% at RPS=256.

\textbf{Challenges.}
IE is carried out by essential co-scaling but builds on indispensable profiling and scheduling. However, building such a serverless DL system with IE support is non-trivial.  
% First, the basic GPU resource quotas for each DL function should be precisely profiled to identify fragments. \textcolor{red}{Diverse factors, including DL function types or priorities}, varying model sizes, and complex execution patterns(e.g., batching for inference), exacerbate the profiling \textcolor{blue}{spaces}\sout{difficulty}\sout{(\textit{Challenge 1})}.
\textbf{\textit{Challenge 1:}} It is costly and difficult to precisely measure the basic GPU resource quotas required to guarantee QoS for each DL function. The profiling result helps prevent overprovisioning and converts previously internal GPU fragments of DL functions into external resources for potential reassignment.
Diverse factors, including DL function types or priorities, varying model sizes, and complex execution patterns (e.g., batching for inference), exacerbate the profiling sampling spaces.
% Second, the real-time workload characteristics of functions and cluster resourcing status expand the search space for scheduling and collocation to achieve multiple goals(e.g., defragmentation, QoS guarantees) (\textit{Challenge 2}).
\textbf{\textit{Challenge 2:}} It requires wise scheduling to reuse GPU fragments and make collocation decisions efficiently. The real-time fluctuations of function workloads and the cluster's resource status expand the scheduling and collocation search space, making it challenging to achieve multiple objectives simultaneously (e.g., defragmentation, QoS guarantees).
% For example, since both the allocated GPU resources of collocated tasks dynamically change, the functions can have implacable resource competition, hurting their QoS or invalidate the scheduling decisions fleetly
% Moreover, it needs a deliberate design to handle resource contention of collocated tasks (\textit{Challenge 3}). For example, under the high RPS (e.g., 256 and 512), there exist serious inference SLO violations in Figure~\ref{fig:toy-inf} and training throughput decrease in Figure~\ref{fig:toy-train}.
\textbf{\textit{Challenge 3:}} A deliberate and cross-layer coordination mechanism is needed to handle resource contention of collocated tasks. For example, under the high RPS (e.g., 256 and 512), there exist serious inference SLO violations in Figure~\ref{fig:toy-inf} and training throughput decrease in Figure~\ref{fig:toy-train}, caused by blunt high-level horizontal scaling and disordered low-level vertical scaling. 

\section{System Design}\label{sec:system-design}

% \textcolor{red}{In this section, we present the design of \projectname{}. We first describe the execution workflow of the entire system in Section~\ref{Sec:design}. Then we provide detailed descriptions of key components. Specifically, Section~\ref{sub-sec:profiling} first profiles resources of each DL instance, including training and inference tasks, to identify available GPU fragments. Second, Section~\ref{sub-sec:scheduling} introduces GPU allocation principles for each instance from the cluster scheduler perspective. Finally, Section~\ref{subsec:scaling} presents the core 2D Co-Scaling mechanism, achieving GPU resourcing-on-demand. } 

This section outlines the design of \projectname{}. We first describe the system's execution workflow (Section~\ref{Sec:design}), followed by detailed explanations of key components. Section~\ref{sub-sec:profiling} profiles resources to identify available GPU fragments. Section~\ref{sub-sec:scheduling} involves GPU allocation principles from the global scheduler’s perspective. Lastly, Section~\ref{subsec:scaling} presents the core 2D co-scaling mechanism for GPU resourcing-on-demand.

\subsection{Architecture Overview}
\label{Sec:design}
\begin{figure}
    \centering
    \setlength{\abovecaptionskip}{2pt} 
    \includegraphics[scale=0.35]{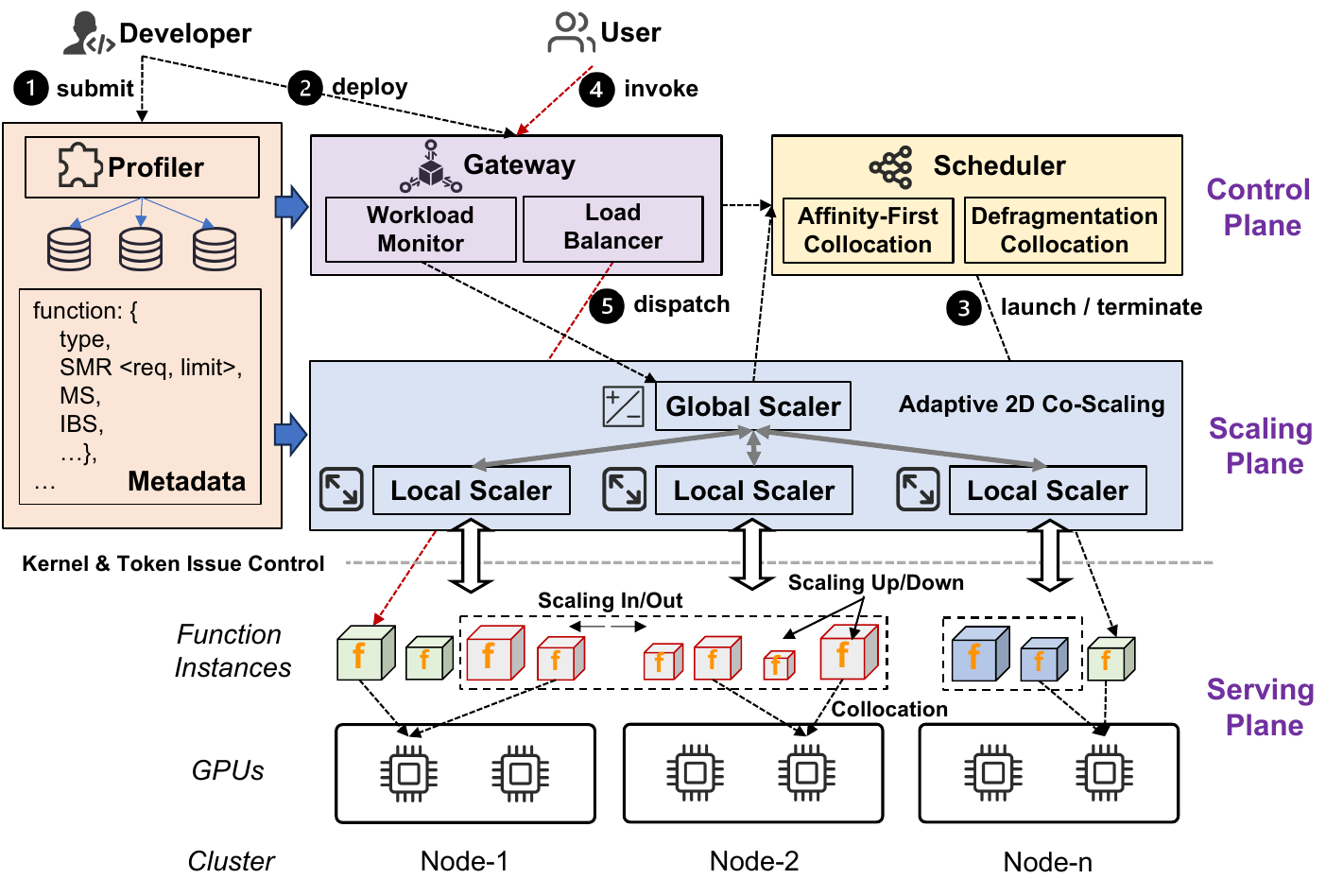}
    \caption{The system architecture of \projectname{}.}
    % \vspace{-0.25in}
    \label{fig:architecture}
\end{figure}

% We build \projectname{}, a Serverless DL system equipped with GPU fragmentation awareness and 2D-elasticity of GPU resource provisioning in response to fluctuating workloads. The system aims to improve GPU utilization, reduce user and provider costs, and enhance the capability of Serverless GPU resource-on-demand across multiple aspects. The system architecture is depicted in Figure~\ref{fig:architecture}. In the control plane, developers submit training or inference codes with specified QoS requirements\footnote{Specifically, we generally consider training throughput (or Job Completion Time) and inference latency as QoS objectives, respectively.}\textcircled{1}. The Profiler rapidly identifies resource configurations that meet the DL task's QoS through pre-running trials and acquires corresponding throughput metadata referred by the Scheduler(Section 3.2). The Scheduler integrate workload affinity and resource complementarity characteristics to deploy new instances \textcircled{3}, maximizing cluster efficiency(Section 3.3).

% We present \projectname{}, a serverless DL system with introspective elasticity support for GPU resourcing-on-demand. It collocates DL functions reasonably with resource-complementarity concerns to reduce GPU fragments, and co-scale resources at the inter-instance and intra-instance level, with high efficiency and QoS guarantees.
\projectname{} is a serverless DL system with introspective elasticity support for GPU resourcing-on-demand. It collocates DL functions with resource-complementarity concerns to reduce GPU fragments, and adaptively adjusts GPU provisioning to guarantee QoS.
The system architecture is depicted in Figure~\ref{fig:architecture}. It consists of a control plane, a scaling plane, and a serving plane. 

\textbf{The Control Plane.} It takes charge of DL task profiling, deploying, scheduling and request dispatching. 
Users submit DL function programs with pre-defined QoS descriptions to the system \ding{182}. Specifically, we generally consider training throughput and inference latency as QoS objectives.
The profiler acquires resource plans with pruning-search trials, as resourcing metadata referred by scheduling.
%TODO(shixiao) and acquires corresponding throughput metadata referred by the Scheduler(Section 3.2). 
After profiling, developers deploy functions to the gateway \ding{183}, which then forwards them to the scheduler.
The scheduler manages instance deployment requirements of DL tasks adhering to several principles. The gateway dispatches all internal and external API requests to the target modules.

\textbf{The Scaling Plane.} It mainly provides an adaptive 2D co-scaling service in horizontal and vertical dimensions, delivering a practical introspective elasticity for GPU resourcing-on-demand.
The global scaler informs the scheduler to carry out horizontal scaling of DL functions, including launching and terminating instances \ding{184}.
The local scaler is distributed in each GPU server. It dynamically adjusts the compute resource of functions by resizing up or down the SM quotas to ensure QoS and improve GPU utilization. Putting them together,
as the external invocation workloads increase \ding{185}, the two-layer scaling will initiate a fast scaling-up and a lazy scaling-out process to deal with bursty workloads and reduce cold starts. Conversely, while the workloads decrease, a fast scaling-down joint with lazy scaling-in will be triggered.

% In the data plane, when user requests are triggered, over the long term, the Scaler makes decisions about scaling in or out based on recent workload, such as Requests Per Second (RPS) (\textcircled{2}). It notifies the Scheduler to create or terminate instances to handle workloads. Considering the significant starting time of DL instances, which cannot instantly (e.g., within 100ms-level) increase serving capacity, Dilu implements a fast scaling-up mechanism to alleviate the issue with QoS guarantees(\textcircled{4}). Additionally, to reduce the cold start frequency and maintain high GPU utilization, Dilu designs a cooperative mechanism of fast scaling down joint with lazy scale-in(Section 3.4), dynamically adjusting SM allocations in real-time to minimize resource fragmentation.

\textbf{The Serving Plane.} The DL functions are running as instances in the serving plane with shared GPUs and other cloud resources, to handle dispatched requests \ding{186}. The functions may employ multiple GPUs or servers for large-scale LLM computations.

\begin{figure*}[t] 
    \centering
    \subfigure[Resnet152]{
            \includegraphics[scale=0.31]{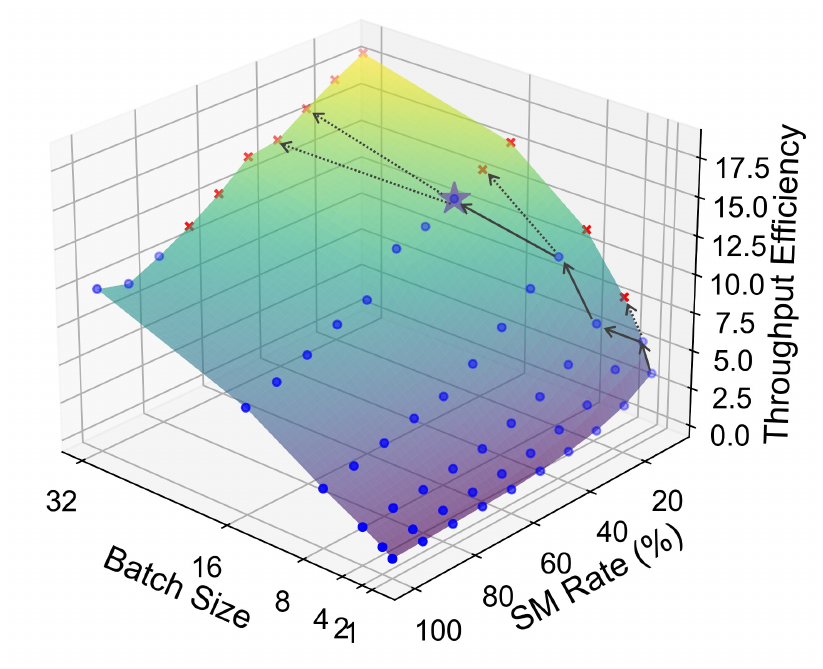}
		\label{fig:mp-resnet}
    }
    \hspace{-0.15in}
    \subfigure[RoBERTa-large]{
        		\includegraphics[scale=0.31]{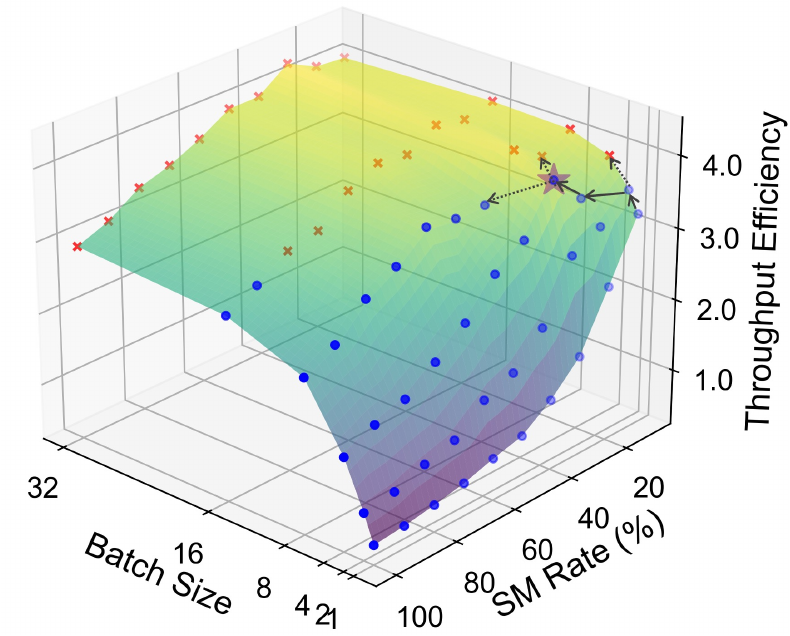}
            \label{fig:mp-roberta}
    }
    \hspace{-0.15in}
    \subfigure[GPT2-large]{
     \includegraphics[scale=0.31]{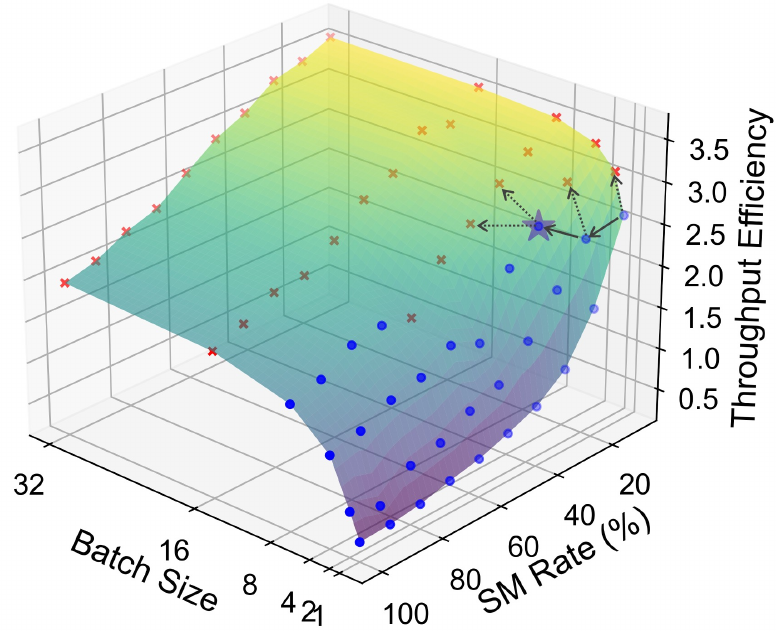}
            \label{fig:mp-gpt2}
    }
    \hspace{-0.15in}
    \subfigure[LLaMA2-7B]{
     	\includegraphics[scale=0.31]{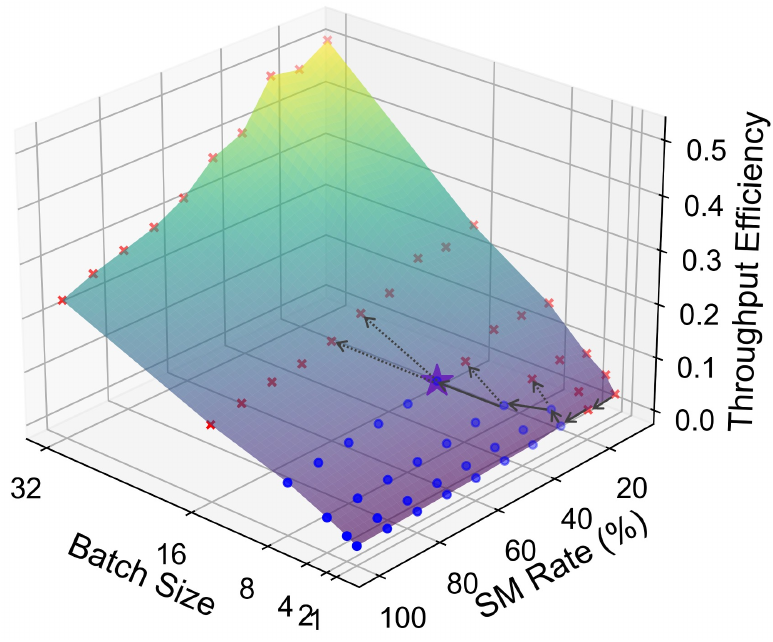}
            \label{fig:mp-llama}
    }
	% \caption{Throughput efficiency performance under varying SM rates and batch sizes for different inference models. With a specific SLO (e.g., 100ms), the star represents the optimal configuration pair <IBS, SMR>. The blue points mean the pairs that SLOs can be met while the red x represents violating SLO. The black solid line represents the forward path and the dashed line represents the blocked path.}
\caption{Throughput efficiency under varying SM rates and batch sizes for inference models with specific SLO targets. The star indicates the optimal configuration pair <IBS, SMR>, blue points denote configurations meeting SLOs, red crosses indicate SLO violations, and the black solid line shows the forward path with the dashed line representing blocked paths.}

	\label{fig:model-profiling}
% \vspace{-0.0in}
\end{figure*}

\subsection{Multi-Factor Profiling}\label{sub-sec:profiling}

% To accurately determine the GPU resource requirements of each DL function while ensuring QoS guarantees, Dilu profiles the execution process to capture detailed needs, including memory size and SM rate(SMR). Commonly, frameworks like PyTorch use a memory pool to minimize the allocation and deallocation of GPU memory, leading to nearly constant memory usage throughout the lifecycle of the function. Thus, the SMR influences training throughput or Job Completion Time (JCT). For inference functions, both batch size(BS) and SMR are critical, impacting the throughput and latency of requests. Consequently, \projectname{} primarily focuses on profiling these three factors.
\projectname{} profiles three key resourcing factors of DL tasks, including the GPU SM rate (SMR), memory size and inference batch size (IBS), for ensuring QoS.
The SMR directly influences training throughput and inference latency. The memory size is usually a constant due to the memory pool management in DL frameworks. The IBS plays a major role as the request batching can improve throughput significantly \cite{infless,batch,choi2022serving,vldb_inf}.

Inspired by the resource quota definition like CPU and host memory in Kubernetes\cite{k8sreqlim}, \projectname{} adopts a similar \textit{<request, limit>} mechanism for SMR quotas, while memory size remains the same due to its steady demand. The \textit{request} quota denotes the minimum compute resource requirement to ensure QoS (e.g., 80\% exclusive training throughput, <100ms inference SLO) and the \textit{limit} indicates the optimal cost-effective quota value, approximately reaching marginal effect points (e.g., near-exclusive training throughput with least compute resource, adaptive for burst inference workloads with SLO ensurance). Leveraging this mechanism, actual allocated SMR can be continuously adjusted between \textit{request} and \textit{limit}, effectively avoiding task starvation and GPU overprovisioning.

\textbf{Training Profiling.} \projectname{} employs a binary search method to iteratively seek the \textit{request} and \textit{limit} quotas of SM. The profiler records exclusive throughput $T_1$ with $high$ (=100\%, firstly) SMR and $T_2$ with $mid$ (=50\%) SMR quota ($low = 0$ indicates zero throughput and is therefore omitted). If $T_2$ is less than $T_1*p$, indicating underprovisioned GPU compute resources, the $low$ value is set to $mid$. Otherwise, the $high$ value is set to $mid$. The profiling ends until the $T_i$ satisfies $T_1*p \pm 2\%$. The SMR for $T_i$ serves as the \textit{request} quota when $p$ is set to 80\% and the \textit{limit} quota when $p=100\%$.

\textbf{Inference Profiling.} We adopt a novel Hybrid Growth Search Strategy to search the most cost-efficient settings of <SMR, IBS>. Although SMR is positively related to throughput, we observe marginal effects when increasing it, as shown in Figure~\ref{fig:model-profiling}.
For example, there is merely a $2\%$ throughput boost for RoBERTa-large model with $\text{IBS}=4$, while increasing $\text{SMR}$ doublely from $50\%$ to $100\%$. 
% Thus, we propose throughput efficacy (TE) defined as $\text{TE} = \frac{\text{Throughput}}{\text{SMR}} = \frac{\text{IBS}}{t_{\text{exec}} \cdot \text{SMR}}$ to express the throughput per SM unit provides.
Thus, we introduce the throughput efficacy (TE) metric, defined as $\text{TE} = \frac{\text{Throughput}}{\text{SMR}} = \frac{\text{IBS}}{t_{\text{exec}} \cdot \text{SMR}}$\footnote{We adhere to \( t_{\text{exec}} =  \text{SLO}/2 \) like INFless \cite{infless}, to account for additional overheads caused by communication, batching waiting, preprocessing, etc.} to denote throughput per SM unit. 
% As illustrated in Figure~\ref{fig:model-profiling}, all the tested models form a convex/concave surface in the three-dimensional space of $\langle \text{IBS}, \text{SMR}, \text{TE} \rangle$. Thus, the HGSS can quickly locate the optimal setting on the surface. In HGSS, the IBS undergoes exponential growth by doubling in each iteration and SMR increments linearly by fixed rate (i.e., 10 units). This approach effectively adapts to the differing sensitivities of the parameters, facilitating a swift convergence to the optimal configuration. Considering some untested models may fail to conform to the expected convex surface, HGSS can still guarantee QoS while the cost-efficiency may be affacted, since the search process is established on the premise of satisfying SLO. We empirically double the profilied SMR \textit{limit} to accommodate bursty workloads with good efficiency. For example, in Figure~\ref{fig:mp-gpt2} of GPT2-large inference with 100ms SLO setting, the black-line path shows the search process with $\langle \text{IBS}=1, \text{SMR}=10 \rangle$, $\langle \text{IBS}=2, \text{SMR}=20 \rangle$, $\langle \text{IBS}=4, \text{SMR}=30 \rangle$, $\langle \text{IBS}=8 \text{SMR}=40 \rangle$ (violates SLO, thus stops). Notably, $\text{BS}=4$ does not mean that each processing batch size is 4; rather, it refers to the maximum size allowed in a batching execution. 
% This hybrid-growth approach effectively facilitates a swift convergence to the optimal configuration.  The profiling reduces overhead significantly while maintaining high accuracy.
As illustrated in Figure~\ref{fig:model-profiling}, all the tested models form a convex surface in the three-dimensional space of $\langle \text{IBS}, \text{SMR}, \text{TE} \rangle$. With this strategy, IBS iteratively increases by doubling during profiling, while SMR increases linearly by a fixed rate (i.e., 10 units). 
While some untested models may not conform to the expected convex surface, the QoS is assured since the search process is established on the premise of meeting SLOs. Finally, the star in each subfigure of Figure~\ref{fig:model-profiling} contains the optimal SMR, marked as the \textit{request} quota.
We empirically set the \textit{limit} quota at twice of \textit{request} to accommodate bursty workloads.

\textbf{Profiling Efficiency}. As Table~\ref{tab:mp-compar} demonstrates, \projectname{} outstands all baselines in search iteration times, 0.7-1.7$\times$ speedup compared with the traversal method, and 1-3.3$\times$ to the SOTA GPUlet \cite{choi2022serving}.
INFless \cite{infless} may sustain lower accuracy due to model decomposition and operator time prediction.

\begin{table}
    \caption{ Inference function profiling comparison for models (a)-(d) as illustrated in Figure~\ref{fig:model-profiling}. The number represents profiling iterations, approximately 30s per trial. 
    }
    % \small
    \centering
    \begin{tabular}{cccccc}
        \toprule
        \textbf{Baseline} & \textbf{a}  & \textbf{b} & \textbf{c} & \textbf{d}  & \textbf{Method}\\
        % & Inception\_V3  & DistillBert & GPT2  & Acc\\
        \midrule
        Traversal              &60  &60  &60  & 60  & pre-running \\
        INFless\cite{infless}  &20  &40  &40   &30  & prediction \\
        GPUlet\cite{choi2022serving} &16  &16  &16   & 16  & pre-running  \\
        \textbf{\projectname{}}     &\textbf{8}  &\textbf{6}  &\textbf{6}   & \textbf{9} & pre-running  \\
        \bottomrule
    \end{tabular}
    
    \label{tab:mp-compar}
    % \vspace{-0.2in}
\end{table}

\subsection{Resourcing-Complementary Scheduling}\label{sub-sec:scheduling}

% The scheduler is responsible for allocating resources and initiating new instances. The primary objective is to improve GPU utilization of cluster GPUs and minimize GPUs occupied, thus maximizing the services provided with limited resources, with guarantees of QoS for each DL function. The objective is formalized as follows: (insert your mathematical formula here, labeled as $\text{Formula XXX}$), with the constraint that the QoS requirements of each task must be met.
% defined by \( t_j(x) \), meets a predefined QoS threshold \( q_j \).
The scheduler manages the GPU allocation plans of DL functions at the cluster level, and makes collocation decisions.
The primary goal is to improve GPU cluster utilization and  
 aggregate throughput, increasing the deployment density \cite{du2022serverless} of DL tasks with QoS guarantees. The objective is formalized in Equation~\ref{eq:objective}, minimizing the number of GPUs used, where $n$ represents the number of GPUs in the cluster, and $g_i=1$ indicates GPU $i$ is occupied, otherwise idle.  Constraint \ref{eq:task-assignment} ensures that each function instance \( f_j \) is allocated to at least one GPU, where $m$ denotes the number of all instances and $f_{ij}=1$ indicates $f_j$ occupies GPU $i$. Constraint \ref{eq:qos} ensures that the execution time of $f_j$ (compute resource requirements) meets the corresponding QoS $Q_j$. Constraint \ref{eq:mem-usage} ensures that the total memory usage of collocated instances remains within the limit of a single GPU card.
 Constraint \ref{eq:gpu-usage} ensures that GPU \( i \) is marked as occupied if it is assigned any \( f_j \). 
% Finally, the binary variables in constraint \ref{eq:binary-variables} specify instance allocations, where \( x_{ij} = 1 \) means instance \( j \) is assigned to GPU \( i \), and \( y_i = 1 \) shows that GPU \( i \) is active.
% \begin{align}
% \textit{min}   \quad & \sum_{i=1}^n y_i \label{eq:objective} \\
% \textit{s.t.} \quad & \sum_{i=1}^n x_{ij} \geq 1, \quad \forall j = 1,  \dots, m, \label{eq:task-assignment} \\
%                         & t_j(x) \geq q_jå, \quad \forall j = 1,  \dots, m, \label{eq:qos} \\
%                         & y_i \geq x_{ij}, \quad \forall i = 1,  \dots, n, \forall j = 1,  \dots, m, \label{eq:gpu-usage} \\
%                         & x_{ij}, y_i \in \{0, 1\}, \quad \forall i = 1,  \dots, n,  \forall j = 1, \dots, m. \label{eq:binary-variables}
% \end{align}

\begin{align}
\textit{min}   \quad & \sum_{i=1}^n g_i \label{eq:objective} \\
\textit{s.t.} \quad & \sum_{i=1}^n f_{ij} \geq 1, \; \forall j = 1,  \dots, m, \label{eq:task-assignment} \\
                        & t(f_j) \leq Q_j, \; \forall j = 1,  \dots, m, \label{eq:qos} \\
                        % & g_i \geq f_{ij}, \quad \forall i = 1,  \dots, n, \forall j = 1,  \dots, m, \label{eq:gpu-usage} \\
                        & \sum_{j=1}^m M(f_{ij}) <=M(g_i),
                        \label{eq:mem-usage} \\
                        \vspace{-0.1in}
                        & g_i = 1 \;\; \text{if} \;\; \sum_{j=1}^m f_{ij} \geq 1 \;\; \text{else}  \;\;g_i=0, \label{eq:gpu-usage} \\
                        & f_{ij} \in \{0, 1\}, \; \forall i = 1,  \dots, n,  \forall j = 1, \dots, m. \label{eq:binary-variables}
\end{align}

% This optimization problem can be regarded as a three-dimensional bin-packing decision problem \cite{}, where one dimension corresponds to memory usage, the second dimension to SM requirement, and the third dimension to the task type. It is well-known as an NP-complete problem, hence we adopt a heuristic greedy algorithm~\ref{algorithm:schedule} to reduce complexity.

The scheduling can be regarded as a 2D bin-packing problem, by coordinating SMR (with IBS) and memory size. It is a well-known NP-complete problem, hence \projectname{} adopts a heuristic greedy Algorithm~\ref{algorithm:schedule} to reduce complexity, which adheres to the following three principles.

\begin{figure}[t]
    \centering
    \subfigure[Without workload affinity]{
        \includegraphics[scale=0.36]{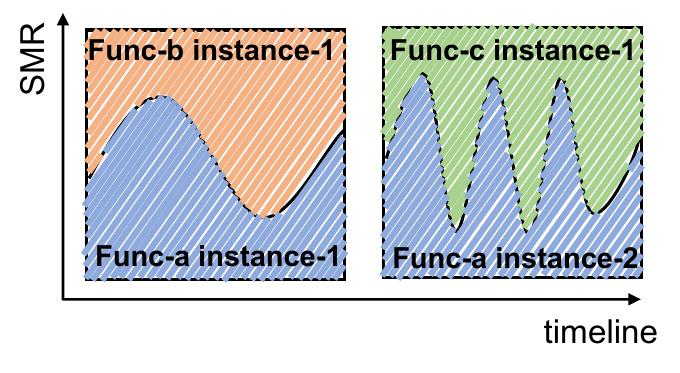}
        \label{fig:workload-unaware}
    }
    \hspace{-0.05\linewidth} % Adjust spacing between subfigures
    \subfigure[With workload affinity]{
        \includegraphics[scale=0.36]{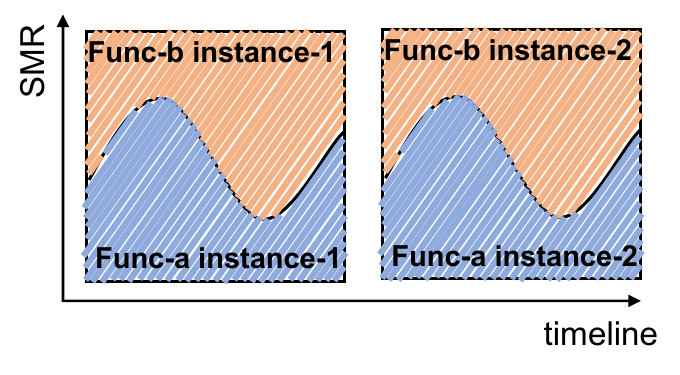}
        \label{fig:workload-aware}
    }
    \captionsetup{skip=-0.1pt}
    \caption{Workload-affinity effect comparison.}
    \label{fig:workloadschedule}
    \vspace{-0.1in}
\end{figure}

\textbf{Principle-1: Reducing laggers with affinity-first collocation.}
Each DL function may differ in its characteristics and workload, and impact the SM consumption accordingly. Random scheduling may lead to situations shown in Figure~\ref{fig:workload-unaware}, where the resources allocated to instances of training Func-a vary differently. It results in a severe barrel effect since the training speed depends on the lagger instance with the least compute resources. Considering the runtime affinity of functions, \projectname{} strategically collocates instances with similar workloads (line 11-12) to reduce the impact of the barrel effect. 
Specifically, instances of the same function (e.g., instance-1/2 of Func-a) often keep analogous loads or SM consumption trends. Thus, as Figure~\ref{fig:workload-aware} shows, \projectname{} tends to collocate the running instance-2 of Func-a with newly launched instance-2 of Func-b, instead of conflicting instances of Func-c. Thus, it mitigates the impact of the barrel effect and prevents elastic scheduling failures.

% \textcolor{red}{Specifically, as Figure~\ref{fig:workload-aware} shows, it tends to launch a new inference instance (i.e., instance-2 of Func-b) on the GPUs where it runs another distributed training instance (i.e., instance-2 of Func-a ) associating to training instance-1 of Func-a, which collocates with the current inference function instance (i.e., instance-1 of Func-b ). Based on the load balancer on the control plane, the requests are evenly dispatched to both instances, avoiding the lagger effect. }
% considering the function affinity realized in running time, \projectname{} prefers to collocate instances with similar workloads (line 11-12), thereby mitigating the impact of the barrel effect. 

% \textbf{Second, differentiating consuming available resources with computing patterns}.
% \textbf{Introspection-2: Resource-complementarity defragmentation principle.} 
\textbf{Principle-2: Defragmentation through Resource Complementarity.} 
% \projectname{} considers both SMR and MS requirements to maximize GPU utilization (corresponding to \textit{SelectOptGPU} in lines 19-29). The \textit{bestScore} represents the corresponding GPU with the minimum weighted fragmentation. With these, the scheduling can search the GPUs with the objective in \ref{eq:objective}. It is noted that for models that can fit within available resources or fragments of a single GPU , we employ a best-fit strategy. For larger models (e.g., LLMs) that exceed available resources of a single GPU, we adopt a worst-fit strategy with multi-GPU to minimize pipeline stages and reduce end-to-end latency. (?????xxxxx)
\projectname{} considers both SM and memory resources to maximize GPU utilization (the \textit{SelectOptGPU} function in lines 19-29). The \textit{bestScore} corresponds to the GPU with the minimum weighted fragmentation. For models that fit within a single GPU fragment, we employ a best-fit strategy. For larger models (e.g., LLMs) that exceed a single GPU fragment, we adopt a memory-based worst-fit strategy which prioritizes choosing GPUs with the most remaining memory, to minimize pipeline stages and reduce end-to-end latency. If no GPU fragments are available, a new GPU instance will be allocated (line 15-16).

% For single-GPU function cases (inference instances????), it employs a best-fit strategy to minimize observed fragmentation. For multi-GPU function cases, a worst-fit strategy (inference instances?????) is adopted to minimize pipeline or other stages, reducing communication load and end-to-end latency. With differentiated resouce matching methods,   (dijin) selectOptGPU and fragmentation relationship.

% \textbf{Introspection-3: QoS guarantee principle.} $\Omega$ and $\gamma$ (line 26) respectively restrict the sum of maximum \textit{request} and \textit{limit} allowed on each GPU. We control these hyperparameters within conservative ranges (i.e. $\Omega$=1,$\gamma$=1.5) to minimize the impact of oversubscription on DL task QoS.

\textbf{Principle-3: Balancing the oversubscription and QoS guarantees.} Higher oversubscription leads to increased function density but will cause severe performance interference. Given the QoS guarantees, two parameters, $\Omega$ and $\gamma$ (line 26), are used respectively to restrict the maximum sum of \textit{request} and \textit{limit} quotas provided per GPU. We conservatively set these hyperparameters (e.g., $\Omega = 1$, $\gamma = 1.5$) to minimize the effects of oversubscription.

\begin{algorithm}[t]

\caption{Heuristic GPU Scheduling Algorithm}
\footnotesize
\begin{algorithmic}[1]
\State \textbf{Input:}
\State \hspace{\algorithmicindent} $G_{act}$: Active GPUs with at least one deployed function.
\State \hspace{\algorithmicindent} $m_j$, $sm_j^{req}$, $sm_j^{lim}$, $n_j$: Resource requirements and number of GPUs needed for function $F_j$.
\State \hspace{\algorithmicindent} $\Omega, \gamma, M_i$: Max allowable sums of SM requests and limits ratios, and memory on each GPU.

\State \textbf{Output:}
\State \hspace{\algorithmicindent} $I^*$: Set of optimal GPUs for deployment.

\Function{ScheduleInstances}{}
    \While{True}
        \State Accept the deployment request for $F_j$ and initialize $I^* \gets \emptyset$.
        \For{$k \gets 1$ \textbf{to} $n_j$}
            \State $G_{WA} \gets \text{GPUs hosting instances with high workload-affinity.} $  
            \State $i^* \gets \Call{SelectOptGPU}{G_{WA}, sm_j^{req}, sm_j^{lim}, \Omega, \gamma}$
            \If{$i^* == -1$}
            \Comment{Select from the GPUs without WA}
                \State $i^* \gets \Call{SelectOptGPU}{G_{act} \setminus G_{WA}, sm_j^{req}, sm_j^{lim}, \Omega, \gamma}$
            \EndIf
            \If{$i^* == -1$}
            \Comment{No available active GPU}
                \State $i^* \gets \text{Start a new GPU instance}$
            \EndIf
            \State Update resource info of $G_{i^*}$ for $sm_j^{req}$ and $sm_j^{lim}$
            \State $I^* \gets I^* \cup \{i^*\}$
        \EndFor
        
        \Return $I^*$
    \EndWhile
\EndFunction

\Function{SelectOptGPU}{$G_{candidates}$, $sm_j^{req}$, $sm_j^{lim}$, $\Omega$, $\gamma$, $m_j$}
    \State $\text{bestScore}, i^*\gets \infty, -1$
    \For{each $i$ in $G_{candidates}$}
        \State $newReqSum \gets \sum_{k \in \text{Funcs on } i} sm_k^{req} + sm_j^{req}$
        \State $newLimSum \gets \sum_{k \in \text{Funcs on } i} sm_k^{lim} + sm_j^{lim}$
        \State $newMemUsage \gets \sum_{k \in \text{Funcs on } i} m_k + m_j$
        \State $score \gets \alpha \cdot \left(1 - \frac{newReqSum}{SM_i^{total}}\right) + \beta \cdot \left(1 - \frac{newMemUsage}{M_i}\right)$
        \If{$newReqSum \leq \Omega$ and $newLimSum \leq \gamma$ and $newMemUsage \leq M_i$ and $score < \text{bestScore}$}
            \State $\text{bestScore} \gets score$
            \State $i^* \gets i$
        \EndIf
    \EndFor
    \State \Return $i^*$
\EndFunction

\end{algorithmic}
\label{algorithm:schedule}
\end{algorithm}

% \begin{figure}[p]
%     \centering
%     \subfigure[Without workload affinity]{
%         \includegraphics[scale=0.36]{figures/workload-unaware.pdf}
%         \label{fig:workload-unaware}
%     }
%     \hspace{-0.05\linewidth} % Adjust spacing between subfigures
%     \subfigure[With workload affinity]{
%         \includegraphics[scale=0.36]{figures/workload-aware.pdf}
%         \label{fig:workload-aware}
%     }
%     \captionsetup{skip=-0.1pt}
%     \caption{Workload-affinity effect comparison.}
%     \label{fig:workloadschedule}
%     % \vspace{-0.2in}
% \end{figure}

\subsection{Adaptive 2D Co-Scaling}\label{subsec:scaling}

% Traditional serverless systems only focus on horizontal elasticity but overlook vertical elasticity, resulting in frequent and costly cold-starts. \projectname{} introduces an \textit{Adaptive 2-Dimension Co-Scaling} (A2DCS) mechanism, which combines fast vertical GPU elastic provisioning (scaling up/down) with lazy horizontal provisioning (scaling in/out). This mechanism not only improves GPU utilization but also better guarantees QoS.

\projectname{} introduces an adaptive two-dimensional co-scaling mechanism, which provides dynamic and fast vertical provisioning (i.e., scaling up/down) elasticity and cost-effective and lazy horizontal elasticity (i.e., scaling in/out). Compared with horizontal-only scaling in classic serverless computing, it strengthens GPU resourcing-on-demand and provides a smooth transition to handle fluctuating workloads, minimizing the impact of cold starts. Specifically, GPU compute provisioning quotas have shifted from traditional discrete integers to continuous decimals (unit is \# of GPU).

% \begin{figure}
%     \centering
%     \setlength{\abovecaptionskip}{1pt} 
%     \includegraphics[scale=0.4]{figures/RCKM.pdf}
%     \caption{The vertical scaling workflow in \projectname{}.}
%     \label{fig:RCKM}
%     % \vspace{-0.3in}
% \end{figure}

\begin{figure}[t]
    \centering
    \setlength{\abovecaptionskip}{1pt} 
    \includegraphics[scale=0.4]{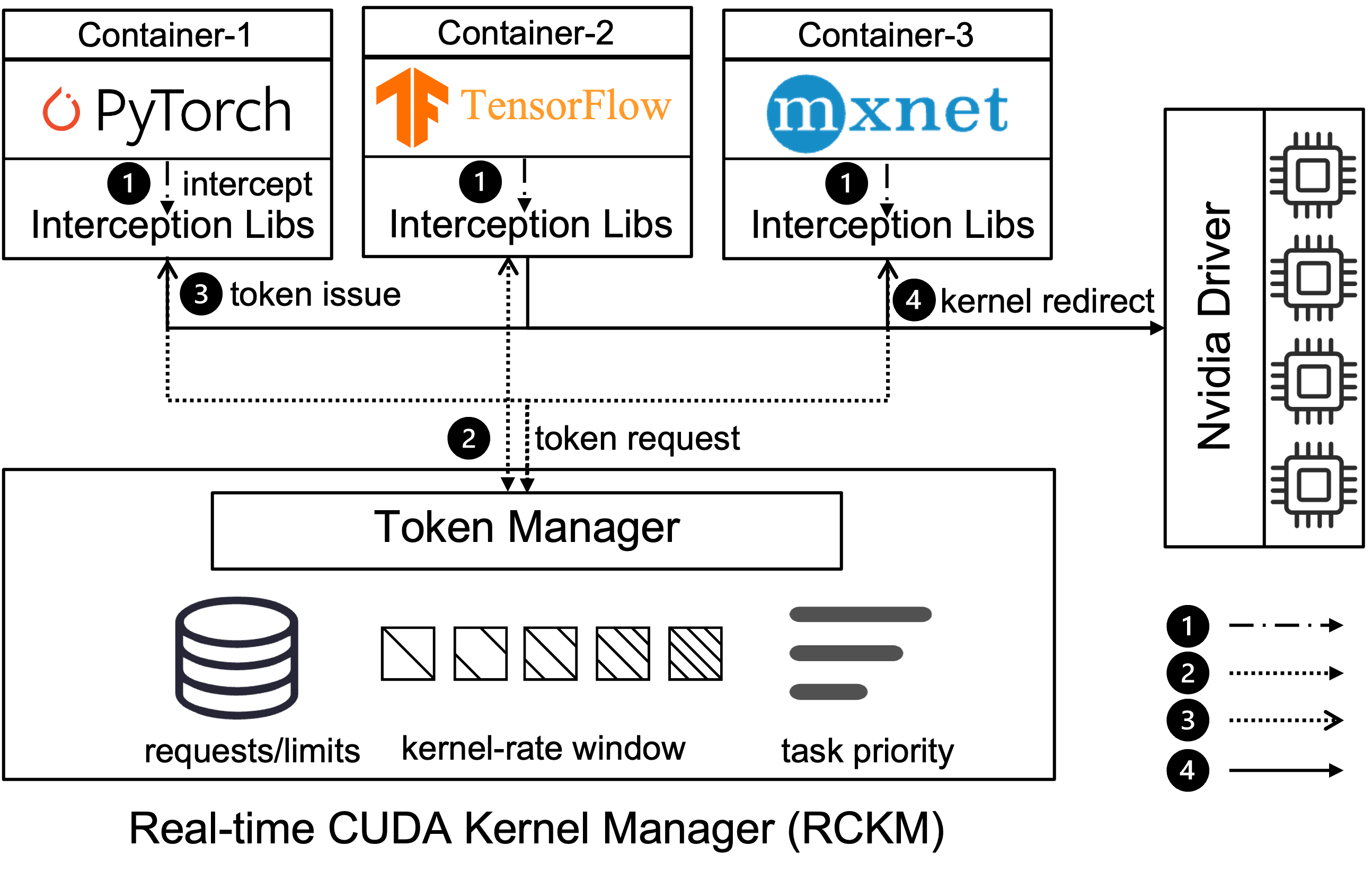}
    \caption{The vertical scaling workflow in \projectname{}.}
    \label{fig:RCKM}
    % \vspace{-0.3in}
\end{figure}

% \subsubsection{Fast/Eager Scaling Up/Down}\label{subsec:fast-scaling}
\subsubsection{Dynamic and Fast Scaling Up/Down} \label{subsec:fast-scaling}
% \projectname{} introduces vertical scaling for two purposes, supporting low-level dynamic GPU sharing and act as a transition for expensive horizontal scaling (in \ref{sec:lazyscalinginout}).

The vertical scaling is responsible for throttling compute resources (i.e., SMs), and has two objectives: to support GPU resourcing-on-demand at the intra-GPU level, and to facilitate a transition to bulky horizontal scaling, as discussed in Section~\ref{sec:lazyscalinginout}.

Since GPU drivers (e.g., NVIDIA) are often closed-source, direct SM management is unfeasible. Inspired by the monitor-and-control mechanisms used in  \cite{tgs,yeh2020kubeshare,gaiagpu,fastg}, \projectname{}  overall adopts a \textbf{server} (Real-time CUDA Kernel Manager, RCKM) \textbf{-client} (Interception Library, IL within each container) architecture to \textit{indirectly} manage SM consumption of each instance, as illustrated in Figure~\ref{fig:RCKM}. They cooperate to monitor and restrict the launched CUDA kernels of collocated instances that shares a single GPU. The workflow mainly consists of the following processes:

\begin{itemize}
    %\item \textbf{Kernel Injection}: CUDA kernels emitted by DL frameworks of each instance are first intercepted by the Interception Libraries (IL) within the container.
    \item \textbf{Kernel Intercept}: CUDA kernel calls emitted by each instance from the host CPU are first intercepted by the Interception Library into respective queues via Linux's LD\_PRELOAD mechanism.
    % \item \textbf{Token Request}: To rapidly forward these kernels, instances ask for tokens from the RCKM every predetermined interval (e.g., 5ms), corresponding to the total kernel numbers triggered during this period.
    \item \textbf{Token  Request}: Token represents available GPU time for each co-located instance per period. To forward kernels, IL first asks for tokens from the RCKM server periodically (e.g., 5ms).
    % \item  \textbf{Token Issue}: RCKM issues token to instances based on requests/limits and task priority metadata obtained from profiling, continuous execution cycles recorded by Token Dispatcher. We discuss it on Algorithm~\ref{algorithm:scalingupdown} in detail.
    %TODO(shixiao): and priority metadata obtained from profiling, profiling
    \item \textbf{Token Issue}: RCKM issues tokens to IL based on several factors, including\textit{<request, limit>} quotas, task priority, and continuous execution cycles. We discuss it on Algorithm~\ref{algorithm:scalingupdown} in detail.
    \item  \textbf{Kernel Redirect}: IL determines to block (\# tokens < current kernel counts) or release (\# tokens >= current kernel counts) CUDA kernels for execution.
\end{itemize}

The Algorithm~\ref{algorithm:scalingupdown} details introspective vertical elasticity, explaining how dynamic GPU provisioning works for each DL instance. The motivation stems from observations via Nsight System \cite{nsys}, where we note that SM contention would prolong the kernel launching cycle (KLC) time within an iteration\footnote{We consider one iteration to encompass a forward and backward propagation for training tasks, as well as a single batch execution for inference tasks.
}, e.g., from 25ms to 50ms for RoBERTa-large inference. Therefore, the core idea is to dynamically allocate GPU tokens for collocated instances on the same GPUs based on KLC changes of high-priority task instances, without monitoring reactive metrics from the application layer. 
% todo

Specifically, when the calculated KLC increases significantly (line 14), it implies either a bursty workload of itself or overly aggressive GPU provisioning for its collocated instances. Then RCKM sets the global variable \textit{state} to $EMERGENCY$ and \textbf{fast} resizes up its issued tokens (line 15). Accordingly, the collocated instances are temporarily resized down based on the KLC variation (line 26-27). Only the current instance can reset or modify the $EMERGENCY$ state. Next, we explain the \textbf{dynamicity}. If no kernels have been launched recently, then the instance will scale down (line 16-17). And if its collocated instances have not launched kernels recently, its quotas will gradually be increased (line 18-19). Otherwise, RCKM will maintain the current provisioning plans in the relatively stable \textit{CONTENTION} state.

% \textcolor{red}{TODO. give a summary about the "what GPU sharing technique is used".}
\textbf{Discussion}. Different from previous spatio-temporal GPU sharing methods in \cite{choi2022serving,fastg,han2024inss}, which all depend on static MPS \cite{NVIDIAMPS} allocation technique, \projectname{} enables basic spatial sharing via collocation and fine-grained token issuing management based on pre-profiled \textit{<request, limit>} meta-quota to avoid task starvation and GPU overprovisioning. In the temporal dimension, it dynamically allocates tokens between \textit{<request, limit>} from a global perspective (RCKM), ensuring high utilization of the entire GPU.
Moreover, Dilu relies on the isolation properties of containers for security, providing reliable protection for multi-tenant services. As for fault tolerance, Dilu leverages the classic restart strategy in serverless.

\subsubsection{Lazy Scaling Out/In}
\label{sec:lazyscalinginout}

\projectname{} also facilitates introspective horizontal scaling, effectively integrating it with vertical scaling to amplify system resilience.

% \projectname{} strengthens horizontal scaling with vertical scaling retrospectively. 
% Classic horizontal scaling (e.g., FaST-GShare \cite{fastg}, INFless \cite{infless}, Azure Serverless  \cite{serverlessinthewild}) launches or terminates function instances \textit{reactively} as workloads increase or drop, integrating with adaptive or prediction-based pre-warming and keep-alive strategies. More or less, they suffer from the cold-start overhead, resulting in slow deployment of DL functions and severe SLO violations. Thus, \projectname{} adopts the fast scaling-up\&down and lazy scaling-out\&in policy. Then, the resourcing adjustment all time first consider to meet the demands by scaling-up\&down current instances. As the resourcing demands keep increasing or decreasing to a level, scaling-out\&in will be activated simultaneously with scaling-up\&down. In this period, both the 2-D scaling are happening, so the cold-start latency of horizontal scaling can be overlapped. The co-scaling provides a continuous resourcing experiences for DL functions. (kankan zheli...........)

Classic horizontal-only scaling (e.g., FaST-GS \cite{fastg}) launches or terminates function instances \textit{reactively} as workloads increase or drop. Advanced approaches (e.g., INFless \cite{infless} and Azure Serverless \cite{serverlessinthewild}) calculate pre-warming and keep-alive duration based on prior knowledge to associate workload prediction with pre-provisioning. However, due to short-term unpredictability, these strategies still incur significant cold start overheads caused by the slow and bulky deployment of large DL functions. It also leads to severe SLO violations. Instead, \projectname{} strengthens horizontal scaling in cooperation with fast vertical scaling to manage the burst workloads.

Specifically, \projectname{} handles sudden workload increases by first triggering fast scaling-up, adaptively delaying the scaling-out execution and avoiding cold starts caused by few-second-level bursty requests. The global scaler maintains a sliding window
(i.e., size=40s) for each function to guide the horizontal scaling. If at least \(\phi_{\text{out}}\) RPS values within the window exceed the serving throughput of deployed instances (acquired from profiling), the scaler informs the scheduler to launch new instances. Here, \(\phi_{\text{out}}\) (i.e., 20) indicates a relatively stable high workload. Conversely, to avoid frequent restarts of new instances, a scaling-in decision is only triggered if more than \(\phi_{\text{in}}\) (i.e., 30) RPS values in the window fall below the serving throughput of (\# of instances - 1). Notably, the lazy scaling-out only occupies a small portion of GPU memory (as shown in Figure~\ref{fig:spatial-feas}). Due to the fast scaling-down mechanism mentioned above, the idle SMs can be dynamically reallocated to other collocated instances.
\begin{algorithm}[t]
\caption{Fast Scale-up/down Control Algorithm}
\footnotesize
\begin{algorithmic}[1]

\State \textbf{Input:}
\State \hspace{\algorithmicindent} $MaxTokens$: Maximum number of tokens that can be issued.
\State \hspace{\algorithmicindent} $RW$: Kernel rate windows for instances.
\State \hspace{\algorithmicindent} $R_{\text{current}}$: Current kernel execution rate.
\State \hspace{\algorithmicindent} $request, limit$: Request/limit rate for the instance.
\State \hspace{\algorithmicindent} $Type$: Type of the instance (e.g., SLO-sensitive).
\State \hspace{\algorithmicindent} $T_{\text{current}}, T_{\text{min}}$: Current and minimum recorded KLCs.

\State \textbf{Output:}
\State \hspace{\algorithmicindent} $R_{\text{issue}}$: Issued tokens of the instance during this cycle.

\Function{IssueToken}{}
    \State Shift Rate Window $RW[current]$ with $R_{\text{current}}$.
    % \State $R_{\text{base}} = MaxTokens * requests$
    % \State $\Delta T = \frac{T_{\text{current}} - T_{\text{min}}}{T_{\text{min}}}$
    \If{Type is SLO-sensitive}
        \State $\Delta T = \frac{T_{\text{current}} - T_{\text{min}}}{T_{\text{min}}}$ \Comment{Calculate the relative change}
        \If{$\Delta T > \eta_{violation}$} \Comment{Trigger protective logic, scale up}
            \State $state, R_{\text{issue}}  \gets EMERGENCY,  MaxTokens * limit $
            % \State $R_{\text{issue}} = MaxTokens * limits$
        \ElsIf{$\textbf{sum}(RW[current]) == 0$} 
        \Comment{Scale down}
            \State $state, R_{\text{issue}} \gets RECOVERY, MaxTokens * request$
        \ElsIf{$\textbf{sum}(RW[others]) == 0 $} \Comment{Scale up}
            \State $state, R_{\text{issue}} \gets RECOVERY, R_{\text{last}}*\eta_{increase}$
        \Else
            \State $state, R_{\text{issue}} \gets CONTENTION, MaxTokens * request$
        \EndIf
    \Else
        % \State $R_{\text{base}} = MaxTokens * request$
        \Switch{$state$}
            \Case{$NONE$}
            \Comment{Without collocation instances}
                \State $R_{\text{issue}} \gets MaxTokens * limit$
            \EndCase

            \Case{$EMERGENCY$}
            \Comment{Scale down}
                \State $R_{\text{issue}} \gets \textbf{min}(MaxTokens * request, R_{\text{last}}) /  \Delta T $
                % \State $R_{\text{issue}} = R_{\text{base}}$
            \EndCase

            \Case{$RECOVERY$} 
            \Comment{Scale up}
                \State $R_{\text{issue}} \gets \textbf{min}(R_{\text{last}} * \eta_{increase}, MaxTokens * limit)$

            \EndCase
            
            \Case{$CONTENTION$}
                \State $R_{\text{issue}} \gets R_{\text{last}}$
            \EndCase
           
        \EndSwitch
    \EndIf
\EndFunction
\end{algorithmic}

\label{algorithm:scalingupdown}
\end{algorithm}
% \vspace{-2pt}

\section{System Implementation}

% \textbf{Prototype system}. We have implemented a prototype system based on Docker and Kubernetes, with 5k lines of code, including 2k+ python LOC for the scheduler and 3k+ C LOC for the scaler to provide 2D-elasticity of GPU provisioning. Additionally, we have developed 3k+ lines of code for simulation, evaluation and scripts. The Profiler, Scheduler, and Horizontal-Scaler are all executed within containers. \projectname{} is compatible with any CUDA-based programming framework like \cite{TensorFlow,PyTorch,mxnet}, offering transparency to developers and end-users. Our code is open-sourced at \textit{XXX}.

\textbf{Prototype system}. We have implemented a prototype system based on Docker and Kubernetes, with 5k lines of code, including 2k+ python LOC for the scheduler and 3k+ C LOC for the scaler. Additionally, we have developed 3k+ lines of code for simulation, evaluation and scripts. The profiler, scheduler, and global scalers are all deployed within containers. \projectname{} is compatible with any CUDA-based programming framework like  \cite{TensorFlow,PyTorch,mxnet}.
% , offering transparency to developers and end-users.

% \textbf{Serverless DL Functions}. A DL function encompasses application layer dependencies such as Pytorch, transformers, and deepspeed, and provides an entry script for training or inference serving. Running on the \textit{NVIDIA Docker} runtime, we equip each DL function deployment with IL and add the IL to the \textit{/etc/ld.so.preload} file.
\textbf{Serverless DL Functions}. In the system, we build a DL function with model-parameter files, an execution entry script and groups of DL runtimes, including Pytorch, transformers, or deepspeed libraries. We also pack IL with each DL function and add its path to the \textit{/etc/ld.so.preload} file.

% \textbf{Profiling}. We have developed a Profiling service based on an efficient binary search algorithm. It accepts DL images and execution code provided by developers, rapidly iterating in a pre-run manner to obtain performance metrics under various configurations. Specifically, the NVIDIA MPS's \textit{CUDA\_MPS\_ACTIVE\_THREAD\_PERCENTAGE} API allocates different SMR sizes to each pre-run instance.
\textbf{Profiler}. It receives DL function images provided by developers. The profiling script varies SMR via \textit{CUDA\_MPS\_ACTIVE- \_THREAD\_PERCENTAGE} of MPS \cite{NVIDIAMPS} API to allocate different compute resources for each pre-running instance.

% , 

% \textbf{Scaler.} The Scaler is divided into a Horizontal Scaler and a Vertical Scaler. The former is located on the cluster master node and operates as a separate service; the latter is present on each node. The Horizontal Scaler monitors workload information for each function obtained from the gateway (including deployment info and the RPS of inference services) and sends scaling-in/out decisions to the Scheduler via HTTP to launch or terminate instances. The Vertical Scaler establishes UNIX connections with the Interception Library (IL) in each GPU DL instance on the managed node, receiving accumulated kernel counts within a period and sending tokens. Notably, both tokens and kernels are measured in units of CUDA Kernel blocks. It is worth mentioning that the IL is bound to the image as a hook library, a transparent process for both developers and end-users.

\textbf{Scaler}. The local vertical scaler on each node establishes a \textit{Unix domain socket} with IL hooked into each instance. It then receives accumulated kernel counts and sends tokens. IL intercepts \textit{cuLaunchKernel},\textit{cuLaunchCooperativeKernel}, and other related APIs. Notably, both tokens and kernels are measured in units of \textit{CUDA kernel blocks}. Each GPU device is managed by a separate \textit{POSIX thread}. The global horizontal scaler periodically (e.g., every 1 second) retrieves workload information from the Application Layer Gateway.

\begin{figure*}[t]
    \centering

    \setlength{\abovecaptionskip}{-0.05cm}
    \subfigure[Inference latency: the dark and light bars resprsent p50/p95, and the mean RPS from left to right is 35, 20, 10 and 3, respectively.]{
            \includegraphics[scale=0.57]{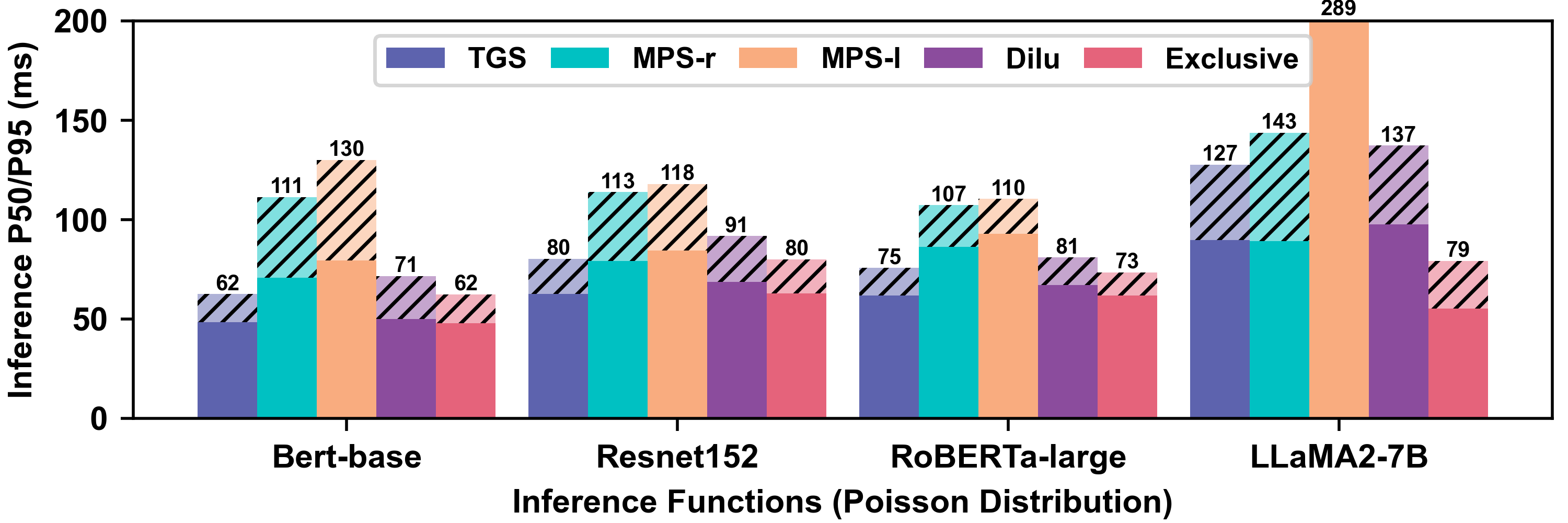}
            \captionsetup{skip=0.02pt}
		\label{fig:train-inf-possion-latency-poisson}
    }
    \hspace{-0.05in}
    \subfigure[Collocated training throughput: error bars represent standard deviation.]{
   		\includegraphics[scale=0.57]{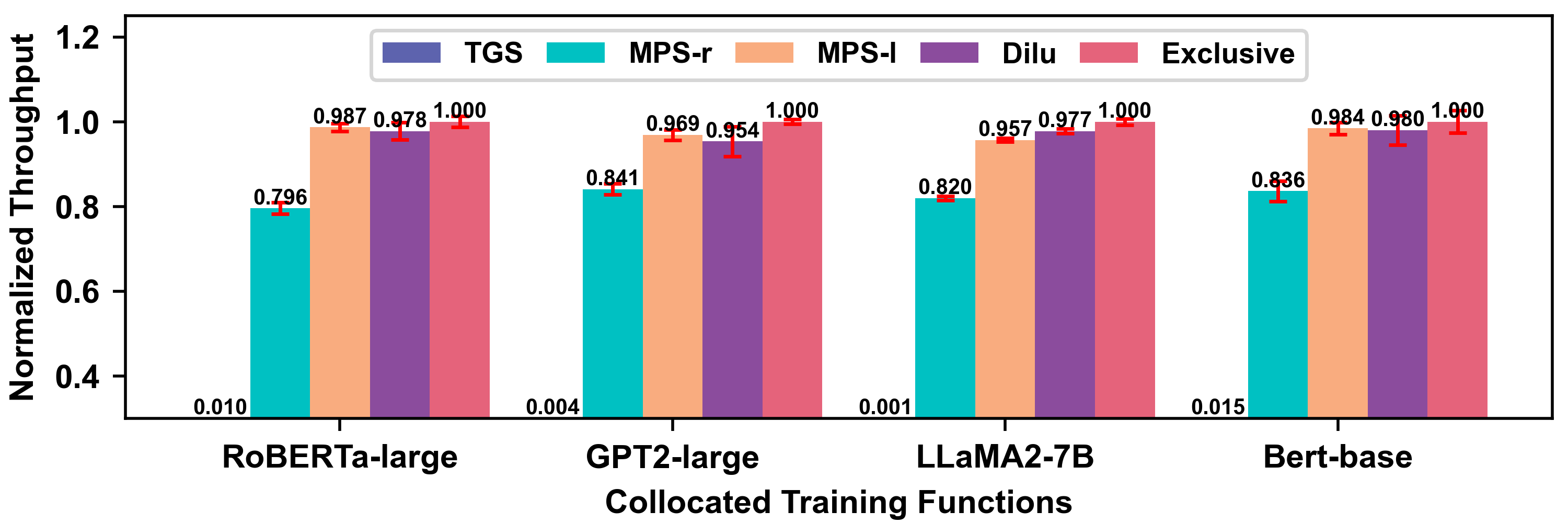}
            \captionsetup{skip=0.02pt}
            \label{fig:train-inf-possion-throughput-poisson}
    }
    \caption{Training-Inference collocation performance.}
    \label{fig:scaling-up/down-train-inf}
% \vspace{-0.05in}
\end{figure*}

\begin{figure*}[t]
    \centering
    \setlength{\abovecaptionskip}{-0.05cm}
    \subfigure[Inference latency under the bursty distribution, and the scaling factor of the initial burst workload from left to right is 4,6,6,4 respectively.]{
            \includegraphics[scale=0.57]{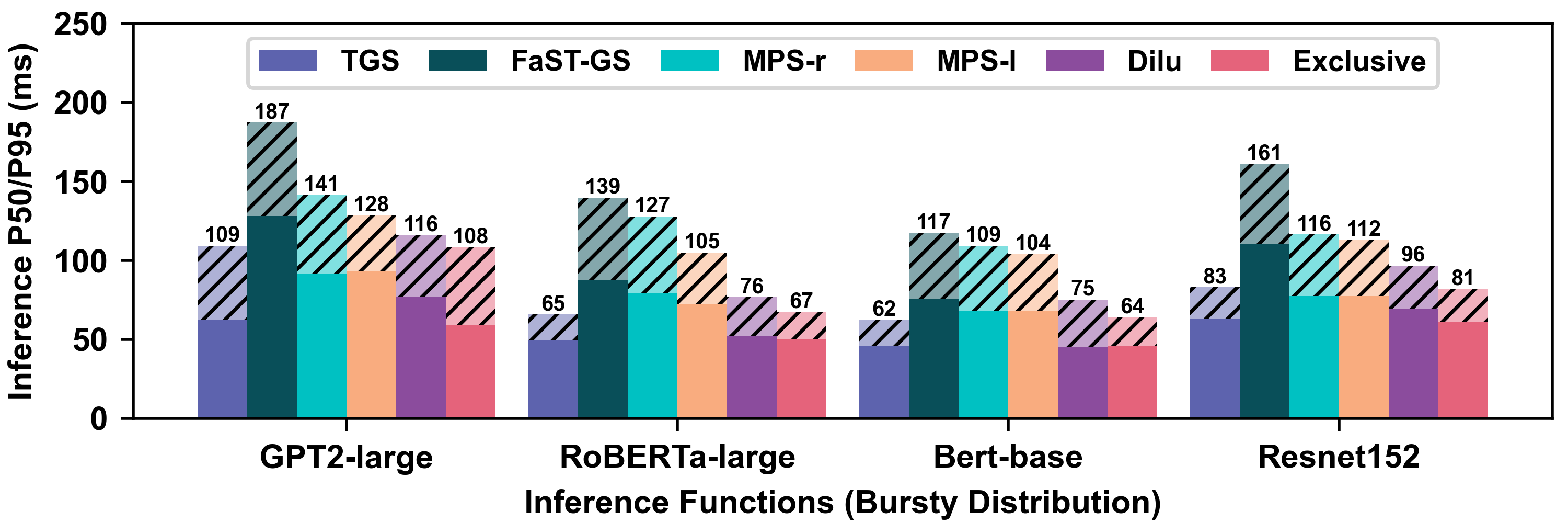}
		\label{fig:inf-inf-possion-latency-bursty}
    }
    \hspace{-0.04in}
    \subfigure[Collocated inference latency under the Poisson distribution, and the mean RPS from left to right is 20, 30, 20 and 3, respectively.]{
   		\includegraphics[scale=0.57]{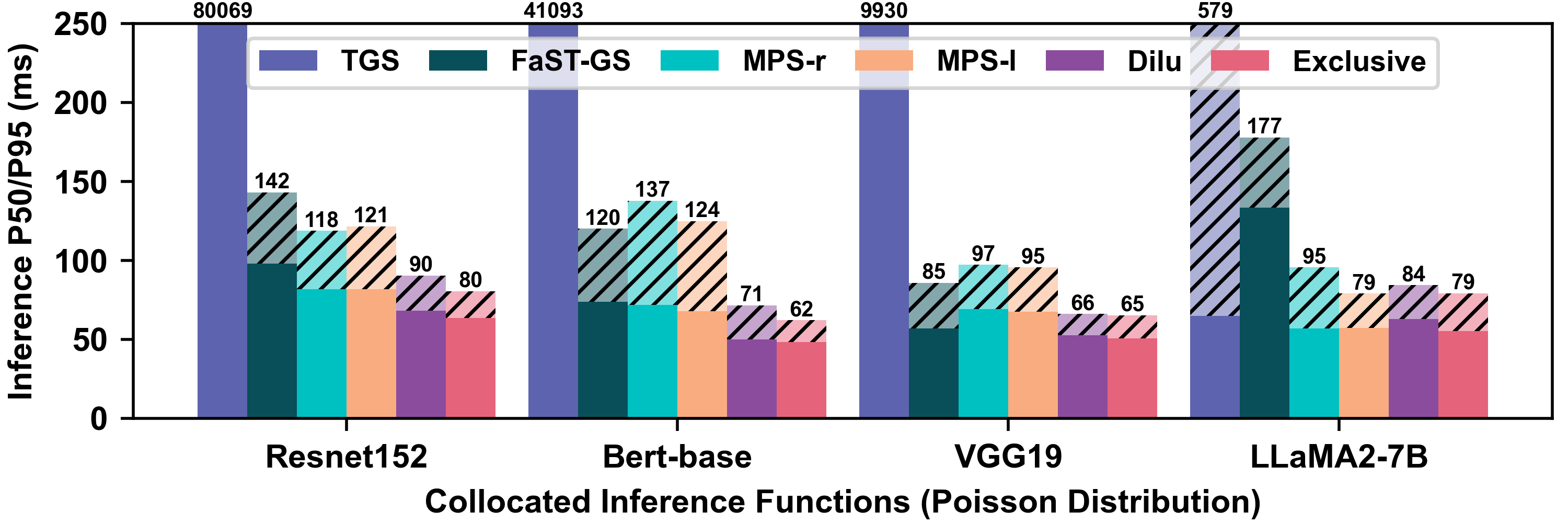}
            \label{fig:inf-inf-possion-latency-poisson}
    }

	\caption{Inference-Inference collocation performance.}
	\label{fig:scaling-up/down-inf-inf}
% \vspace{-0.05in}
\end{figure*}

\textbf{Scheduler}. The scheduler accesses the resource requirement metadata of each function from the profiler and allocates \textit{ip\_address, GPU index, port} for launched instances. For distributed DL function instances, we leverage \textit{NCCL \cite{nccl}} to communicate and adopt the \textit{accelerate \cite{accelerate}} library for the LLM model partition. All instances run on the \textit{NVIDIA Docker} runtime.

\section{Evaluation}\label{sec:evaluation}
\subsection{Methodology}

\label{sec:methodology}
% \textbf{Experiment testbed}. We build a 5-machine Kubernetes(1.23.4) local cluster, each equipped with 48 Intel Xeon(R) Platinum 8163 CPUs, 4 NVIDIA A100-40GB-PCIE GPUs, 256GB RAM, and CentOS Linux release 7.9.2009. The software environments mainly include NVIDIA driver 515.105.01, CUDA 11.7, Docker 24.0.5, PyTorch 1.11, DeepSpeed 0.11.1, and NCCL 2.10.3.

\textbf{Experiment testbed}. We set up an environment of a self-hosted, five-worker Kubernetes (v1.23.4) cluster, each equipped with 4 NVIDIA A100-40GB-PCIE GPUs. Each node is based on PyTorch v1.11, DeepSpeed v0.11.1, NCCL v2.10.3, NVIDIA driver v515.105.01, CUDA v11.7 and Docker v24.0.5.

% \textbf{Workloads}. For the models used in \projectname{}, we consider famous DNN models from computer vision (CV, including ResNet152\cite{he2016deep}, VGG19\cite{vgg19}), natural language processing (NLP, including BERT-base\cite{2019-bert}, RoBERTa-large\cite{liu2019roberta}, GPT-2 large\cite{gpt2large}), and large language models (LLMs) (including LLAMA2-7B\cite{llama2}, ChatGLM3-6B\cite{chatglm3}), with sizes ranging from 0.2GB to 12.6GB. For training functions, we adopt the torch.DDP for medium- and small-size models, and DeepSpeed pipeline-parallelism\cite{DeepSpeed} to fine-tune LLMs such as LLaMA2-7B.

\textbf{Workloads}.
Several popular DL models are selected for evaluation, including ResNet152 \cite{he2016deep}, VGG19 \cite{vgg19} from computer vision, and BERT-base \cite{2019-bert}, RoBERTa-large \cite{liu2019roberta}, GPT2-large \cite{gpt2large} from natural language processing, and LLAMA2-7B \cite{llama2}, ChatGLM3-6B \cite{chatglm3} of the thriving LLM family. The model parameters range from 0.2GB to 12.6GB.

For training, we adopt the torch.DDP \cite{DDP} for medium- and small-size models, and DeepSpeed pipeline-parallelism \cite{DeepSpeed} for LLM fine-tuning. 
For inference, several workload patterns are considered, including Poisson distribution (used by  \cite{orion,batch,vldb_inf, zhong2024distserve}), Gamma distribution (used by FastServe \cite{wu2023fast}), and three typical traces from Azure Function's Production Traces \cite{serverlessinthewild}: Bursty, Sporadic, and Periodic (used by INFless \cite{infless}).

To study the large-scale performance (Section \ref{subsec:simu}), we simulate a cluster of 1,000 nodes, each equipped with 4 GPUs. We generate 3,200 DL instances of varying types in the cluster, with the distribution of training, LLM inference, and non-LLM inference instances by a ratio of 2:2:6.

% \textbf{Metrics}. We use normalized throughputs (e.g. images/s of CV, tokens/s of NLP) relative to the Exclusive baseline for training and p50/p95 latency (ms) or SLO violation rate (SVR)for inference, as the evaluation metrics. For the LLM inference, we record the average time-per-output-token (TPOT) latency of requests.
\textbf{Metrics.} Training throughput (e.g. images/s of computer vision models, tokens/s of natural language processing models), latency (e.g., p50/p95) and SLO violation rate (SVR), cold start count (CSC) of inferences are measured. For the LLM inferences, the average time-per-output-token of requests is regarded as LLM latency.
 
% \textbf{Baselines}. For GPU sharing on vertical scaling,  we compare \projectname{} with the following methods. 
\textbf{Baselines}. They are chosen for both GPU- and cluster-level comparisons. At the GPU level, the baselines include: 
\begin{itemize}
    \item \textbf{Exclusive}: All GPUs are allocated exclusively to DL function instances via pass-through.

    \item \textbf{NVIDIA MPS} \cite{NVIDIAMPS}: The official NVIDIA GPU sharing mechanism is widely used in serverless DL systems \cite{infless,fastg,dhakal2020gslice}. \textbf{MPS-l} (MPS with \textit{limit} quotas from \projectname{} profiling) and \textbf{MPS-r} (MPS with \textit{request} quotas from \projectname{} profiling) are used for comparisons.

    \item \textbf{FaST-GS} \cite{fastg}: A typical spatio-temporal GPU sharing method, specifically designed for serverless DL inference, relies on MPS. For fairness, we allocate the same amount of SMR spatially as MPS-l.

    \item  \textbf{TGS} \cite{tgs}: A transparent GPU sharing method, improving opportunistic job throughput while guaranteeing productive jobs.
\end{itemize}

% \begin{itemize}[leftmargin=*,itemsep=0pt,parsep=1pt,topsep=1pt,partopsep=1pt]
%     \item  \textbf{Exclusive.} All GPUs are allocated exclusivley to DL function instances via pass-through.
%     \item  \textbf{NVIDIA MPS}\cite{infless}. NVIDIA official GPU sharing mechanism, widely used in serverless DL systems\cite{infless,fastg,dhakal2020gslice}. % Due to its static allocation, we have divided it into two versions, namely \textbf{MPS-l} (limits) and \textbf{MPS-r} (requests), representing GPU over-provisioning and under-provisioning, respectively.
%     \textbf{MPS-l} (MPS with SMR limit quotas from \projectname{} profiling) and \textbf{MPS-r} (MPS with SMR request quotas from \projectname{} profiling) are used for cases, representing GPU lower bounded-provisioning and under-provisioning situations respectively.
%     % representing the GPU provisioning under conditions of sufficient and less-than-sufficient resources, respectively.
%     \item  \textbf{TGS}\cite{tgs}. A transparent GPU sharing method, improving opportunistic training job throughput while guaranteeing productive jobs. % It only supports two collocated DL instances on a single GPU.
% \end{itemize}
% For cluster scheduling and horizontal scaling, the baselines are as follows.

At the cluster level, the baselines include:
\begin{itemize}
    \item \textbf{Exclusive}: It is used by \cite{ACI,elasticflow,mlsyssla,hydrozoa} to allocate GPUs exclusively for DL functions, which is a common basic scheme in Kubernetes.
     \item \textbf{FaST-GS+} \cite{fastg}, \textbf{INFless+} \cite{tgs}: Two serverless inference systems based on MPS improve serving throughputs. We extend them to support training scheduling as FaST-GS+ and INFless+.
    
\end{itemize}

% \begin{itemize}[leftmargin=*,itemsep=0pt,parsep=1pt,topsep=1pt,partopsep=1pt]
%     \item  \textbf{Kubernetes-exclusive.} It is used by\cite{ACI,elasticflow,mlsyssla,hydrozoa} to allocate GPUs exclusively for DL functions, a common basic scheme.
%     \item  \textbf{FaST-GShare+}\cite{fastg},\textbf{INFless+}\cite{tgs}. Two serverless inference systems based on MPS improve serving throughputs. We extend them to support training as FaST-GShare+ and INFless+.
%     % \item  \textbf{INFless+}\cite{tgs}. INFless is a serverless native inference system using MPS to consume partitioned GPU compute resources. We extend it to support trainings as INFless+.
% \end{itemize}

% \subsection{Local Cluster Evaluation}

% \subsubsection{Vertical Scaling}

\subsection{Vertical Scaling Performance}
\label{sec:vertical-scaling}
\textbf{High GPU utilization and aggregate throughput}. We analyze the vertical scaling performance in \projectname{} on three typical collocation cases. Experiments show that \projectname{} effectively achieves GPU resourcing-on-demand, dynamically adjusting resource provisioning while ensuring QoS. \projectname{} outperforms all the GPU sharing baselines, and is close to the Exclusive mode which occupies more GPUs and generates an amount of resource fragmentation.

% As mentioned in Section~\ref{sec:methodology}, we evaluate vertical scaling/elasticity through three cases: \textit{training-inference}, \textit{inference-inference}, and \textit{training-training}. Next, XXX

\textbf{Training-inference collocation}.
% As shown in Figure~\ref{fig:train-inf-possion-latency-poisson}, the vertical scaling can balance the resource requirements of the heterogeneous collocated DL tasks.
% Figure~\ref{fig:train-inf-possion-latency-poisson} shows the average p50 and p95 inference latencies under the Poisson distribution and Figure~\ref{fig:train-inf-possion-throughput-poisson} represents normalized average throughput and its standard deviation (std) of coloated training tasks.
As shown in Figure~\ref{fig:scaling-up/down-train-inf}, compared to the Exclusive, \projectname{} achieves only $1.24\times$ p50 and $1.28\times$ p95 latencies on average along with 97.2\% Exclusive' throughput, while saving 50\% of GPU resources.
TGS shows similar inference performance and sub-optimal p95 latency of all, but it nearly stops the collocated training functions. Because TGS simply prioritizes the high-priority inference instance to execute first, slowly and incrementally increases execution opportunities for another low-priority instance through trial. It may not lead to starvation, but it lacks the mechanism of adapting to highly hybrid and fluctuated DL workloads.
MPS-r results in higher inference tail latencies and lower training throughput due to its static and conservative resource provisioning. Compared to the average p50 and p95 latencies of MPS-l and MPS-r, \projectname{} obtains reductions of 35\% and 25\%, and 13\% and 21\%, respectively.
Especially, the LLaMA2-7B inference instance is deployed using four fragmented GPUs, except the Exclusive baseline. We can see that \projectname{} balances training throughput and inference latency, achieving the highest aggregate performance.

% Figure~\ref{fig:train-inf-possion-latency-poisson} shows the p50 (dark color) and p95 (light color with diagonal stripes) latencies of inference instances under Poisson distribution workloads, while Figure~\ref{fig:train-inf-possion-throughput-poisson} displays the normalized throughput for training instances co-located with each inference instance. The error bars of Figure~\ref{fig:train-inf-possion-throughput-poisson} show the standard deviation (std) of training throughputs in the 5-iteration level.
% Among these, LLaMA2-7B inference performance is across-GPU deployed based on resource fragments of 4 GPUs, except in the Exclusive baseline where it can hold the whole model independently. Overall, the Exclusive performs the best due to the absence of resource contention, albeit occupying two GPUs. TGS achieves sub-optimal p95 latency but at the expense of significantly reduced training throughput, due to its inherent time-share mechanism. MPS-r results in higher inference tail latencies and lower training throughput due to its conservative resource provisioning. MPS-l leads to the highest p95, as it intensely contents SMs with the co-located training instance (for fairness, even when the sum of allocated SMR resources exceeds 1, they are still placed on a single GPU). 

\textbf{Inference-inference collocation}.

% Figure~\ref{fig:inf-inf-possion-latency-bursty} demonstrates the inference latency following a bursty distribution, while Figure~\ref{fig:inf-inf-possion-latency-poisson} shows co-located inference instances under a Poisson distribution. It is important to note that TGS's sharing mechanism requires at least one low-priority and one high-priority task process. In this case, we designate the inference instances in the bursty distribution (left figure) as high-priority, and those on the right as low-priority. For the left-side inference tasks, the overall trend aligns with Figure~\ref{fig:train-inf-possion-latency-poisson}, with the only difference being that MPS-l performs better than MPS-r due to reduced resource contention. However, for inference instances co-located with the models in the left figure, TGS's average p50/p95 are 442$\times$ and 405$\times$ that of Dilu, respectively. Furthermore, compared to MPS-l, Dilu reduces the average P95 by 25\% due to RCKM's capability to handle burst loads. In the right figure, where LLaMA2-7B is co-located with Resnet152, Dilu's p95 is slightly higher than MPS-l, largely due to RCKM issuing relatively conservative tokens to LLaMA2-7B inference instances to guarantee the overall performance of Resnet152. In sum, for the \textit{inference-inference} case, \projectname{}performs the best, second only to the Exclusive.
In this case, \projectname{} still keeps the best performance and consistently attains higher GPU utilization than other baselines.
% Figure~\ref{fig:inf-inf-possion-latency-bursty} and Figure~\ref{fig:inf-inf-possion-latency-poisson} show collocated inference latencies under bursty distributions and Poisson distributions, respectively. 
The inference performance in Figure~\ref{fig:inf-inf-possion-latency-bursty} almost aligns with Figure~\ref{fig:train-inf-possion-latency-poisson}, with the only difference that MPS-l performs better than MPS-r due to less resource contention in current collocation mode.
% Since TGS requires to assign different priorities to tasks explicitly, we designate the inference instances for the bursty workloads (left) as high-priority and the other one (right) as low-priority.
In Figure~\ref{fig:inf-inf-possion-latency-poisson}, TGS's average p50 and p95 latencies are 442$\times$ and 405$\times$ of \projectname{} respectively, due to its conservative and speculative sharing mechanism previously mentioned.
Compared to MPS-l in Figure~\ref{fig:inf-inf-possion-latency-poisson}, \projectname{} reduces the average p95 by 25\% through fast vertical scaling to manage bursty workloads. In Figure~\ref{fig:inf-inf-possion-latency-poisson}, the p95 latency of LLaMA2-7B inference with \projectname{} is slightly higher (less than 6\%) than MPS-l, due to the relatively fair token issuing to ensure SLOs of both collocated instances. Additionally, we include comparisons with FaST-GS, a spatio-temporal sharing mechanism. Since it relies on MPS, its performance upper limit matches that of MPS-l only when compute resources are not temporally shared. However, frequent collection of CUDA Event time statistics and the prioritized dequeuing mechanism for temporal sharing introduce significant overhead, leading to higher latency than MPS-l in most cases, as shown by the green bars. This gap becomes negligible for smaller models, such as Bert-base and VGG19.

% However, frequent CUDA Event time statistics and the prioritized dequeuing mechanism designed for temporal sharing yield a significant amount of overhead, making the latency higher than MPS-l in most cases, as indicated by green bars. The gap becomes marginal when the models are small, like Bert-base and VGG19. 

\begin{figure}[t]
    \centering
    \includegraphics[scale=0.56]{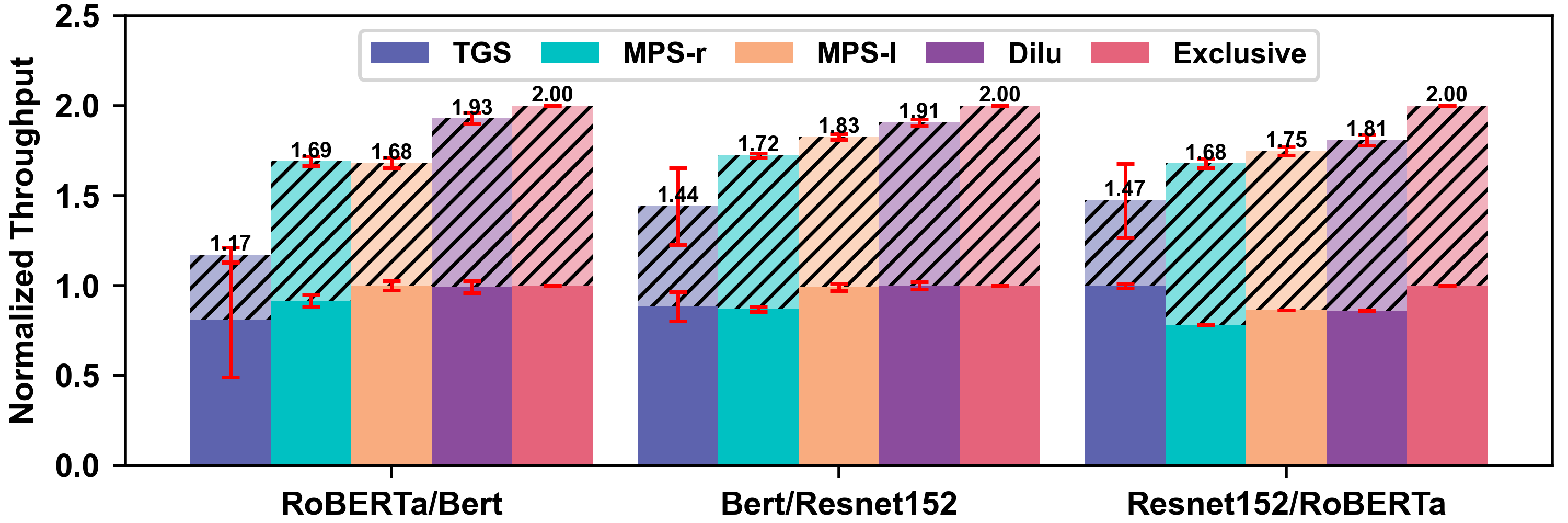}
    % \captionsetup{skip=0.05pt}
    \caption{Training-Training collocation performance: the bottom bars represent the left models.}
    \label{fig:train-train-throughput}
    \vspace{-0.1in}
\end{figure}

\textbf{Training-training collocation}.
% Due to gradients and activations generated during training iterations occupying substantial amounts of memory, we reduce the batch size of each training task to adapt to collocation without memory swapping. Figure~\ref{fig:train-train-throughput} presents the performance comparison results, (todo to caption?)where the lighter bars represent training function-1 and the darker bars with diagonal stripes represent training function-2. The error bars indicate the throughput variance over five iteration levels. The bar height represents aggregate throughput. We observe that Dilu, on average, experiences less than a 12\% loss in aggregate throughput relative to the Exclusive but saves 50\% on GPU costs, outperforming other baselines. In the instances of RoBERTa-large and Bert-base, intense contention due to SM oversubscription results in even lower aggregate throughput for MPS-l than for MPS-r. TGS, with its conservative time-share mechanism, can only guarantee one high-priority training task, while the low-priority training throughput significantly decreases, as indicated by the larger standard deviation.
On average, \projectname{} achieves 176\% of the aggregate training throughput of Exclusive, outperforming all other baselines.
% Since the gradients and activations for training iterations occupy substantial GPU memories, we reduce the batch size of each training task to avoid memory swapping for the collocation. 
As shown in Figure~\ref{fig:train-train-throughput}, \projectname{} is 10\%-14\% and 3\%-14\% higher than MPS-l and MPS-r. We also observe intense SM contention in the RoBERTa and BERT collocation case, resulting in lower throughput for MPS-l compared to MPS-r. TGS still prioritizes high-priority tasks, but the performance of collocated low-priority tasks is severely affected.

% We also observe that there is intense SM contention in RoBERTa and Bert collocation case, leading to a lower throughput of MPS-l than MPS-r.
% TGS still preferentially guarantees high-priority tasks, while the collocated low-priority tasks are damaged severely.
% as indicated by the larger standard deviation and shorter bars.

\begin{figure}
    \centering
    \subfigure[RoBERTa-large with  RPS=64]{
        \includegraphics[scale=0.53]{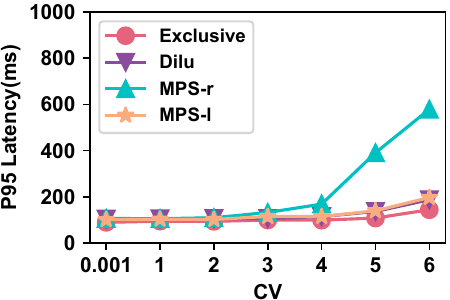}
        \label{fig:p95_roberta_varying_CV}
    }
    \hspace{-0.015\linewidth} % Adjust spacing between subfigures
    \subfigure[GPT2-large with RPS=48]{
        \includegraphics[scale=0.53]{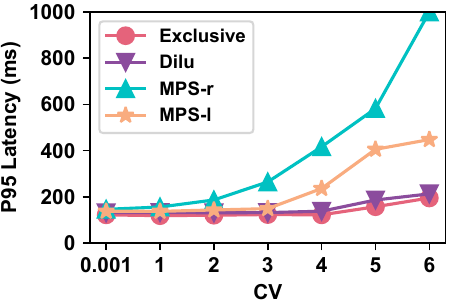}
        \label{fig:p95_gpt2_large_varying_CV}
    }
    % \captionsetup{skip=-0.0pt}
    \caption{Inference latency under gamma distributions, collocated with Bert-base and RoBERTa-large training instances, respectively.}
    \label{fig:scaling-up/down-diff-CV}
    % \vspace{-0.05in}
\end{figure}

% \textbf{Adative adjustment capability}. 
\textbf{Fast adaptivity.}
Under fluctuating Gamma distribution workloads, \projectname{} guarantees SLOs like the Exclusive setup and surpasses other baselines, thanks to the fast scaling-up capability of RCKM. The p95 latencies of \projectname{} on the smaller RoBERTa-large model are only 6\%-29\% higher than the Exclusive in Figure~\ref{fig:p95_roberta_varying_CV}, and 7\%-9\% higher on the bigger GPT2-large model in  Figure~\ref{fig:p95_gpt2_large_varying_CV}.
However, MPS-l and MPS-r both show an exponential growth trend as the Coefficients of Variation (CV) increase due to static resource provisioning. 
Specifically, at CV=6, the p95 of MPS-l and MPS-r are 2.08$\times$ and 4.76$\times$ higher respectively than \projectname{}, while Dilu is only 9\% higher than Exclusive. 
% This is owing to RCKM's protective mechanism under extremely bursty inference workloads, adaptively reducing the tokens issued to the collocated ones.
% For continuous and stable high workloads, the horizontal scaler will launch new instances to avoid performance interference with collocated instances.

\begin{figure}
    \centering
    \subfigure[Training Overhead]{
        \includegraphics[scale=0.53]{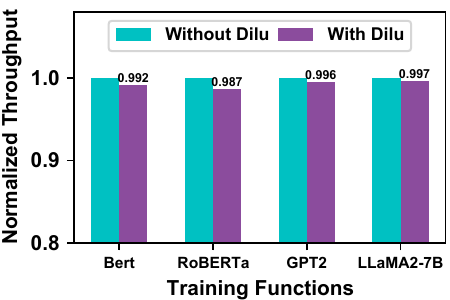}
        \label{fig:train-breakdown}
    }
    \hspace{-0.015\linewidth} % Adjust spacing 
    \subfigure[Inference Overhead]{
        \includegraphics[scale=0.53]{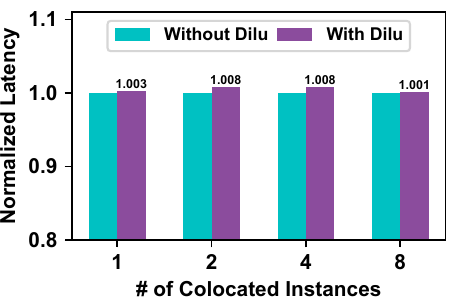}
        \label{fig:inf-breakdown}
    }
    \captionsetup{skip=-0.0pt}
    \caption{Vertical scaling overhead.}
    \label{fig:node-overhead}
    % \vspace{-0.05in}
\end{figure}

\begin{figure*}
    \centering
    \setlength{\abovecaptionskip}{-0.cm}
    \subfigure[Case-1: low inference workloads]{
            \includegraphics[scale=0.57]{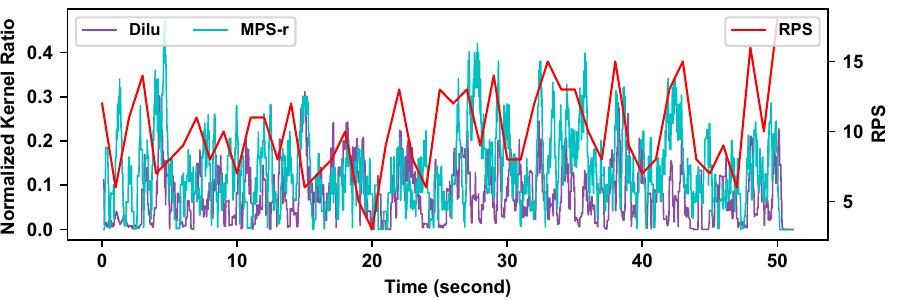}
		\label{fig:kernel_counts_ratio_training}
    }
    \hspace{-0.04in}
    \subfigure[Case-2: fluctuating inference workloads]{
   		\includegraphics[scale=0.57]{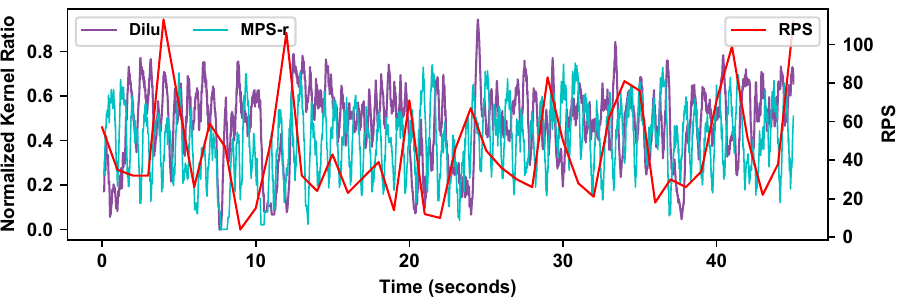}
            \label{fig:kernel_counts_ratio_inference}
    }

	\caption{Kernel issuing traces analysis. The normalized kernel ratio denotes inference kernel counts divided by the total kernel counts of all collocated tasks.}
	\label{fig:scaling-up/kernel-counts}
% \vspace{-0.15in}
\end{figure*}

\begin{figure}
    \centering
    \includegraphics[scale=0.58]{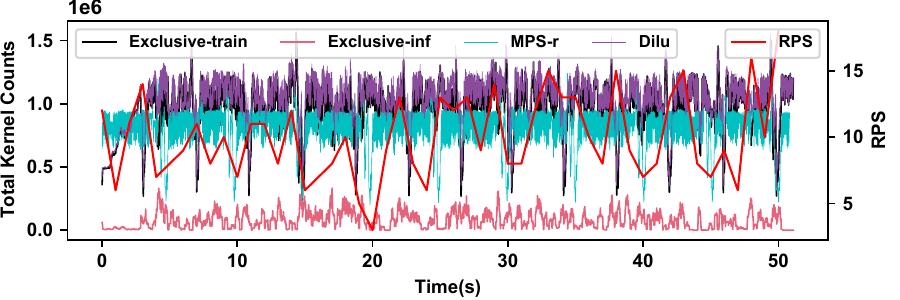}
    % \captionsetup{skip=-0.05pt}
    \caption{Total kernel counts comparison. }
    \label{fig:train-inf-total-kernel-counts}
    % \vspace{-0.05in}
\end{figure}

% \textbf{Case Study.} 
% For further exploration, we record 250ms-level normalized kernel rate timelines in two previous cases to reflect the superiority of vertical elasticity. Figure~\ref{fig:kernel_counts_ratio_training} is based on <training: LLAMA2-7B, inference: RoBERTa-large> collocation depicted in Figure~\ref{fig:scaling-up/down-train-inf}. During this period, as the RPS is at a relatively low level (mean of 10 reqs/s), RCKM increases the tokens allocated to the training instance, unlike MPS-r which keeps a fixed and conservative rate, resulting in an overall training throughput decrease of 15\% than Dilu. Figure~\ref{fig:kernel_counts_ratio_inference} shows the case of Figure~\ref{fig:p95_gpt2_large_varying_CV} at CV=5, showing that under fluctuating workloads, Dilu's line is almost always above MPS-r, explaining why the p95 of MPS-r is 2.1$\times$ that of Dilu.
\textbf{Adaptive kernel issuing}. In Figure~\ref{fig:scaling-up/kernel-counts}, we explore kernel issuing traces of <RoBERTa-large, LLaMA2-7B> case in Figure~\ref{fig:scaling-up/down-train-inf} and <GPT2-large, RoBERTa-large> case when CV = 5 in Figure~\ref{fig:p95_gpt2_large_varying_CV}. 
When the workload is low (average 10 req/s, as shown in Figure~\ref{fig:kernel_counts_ratio_training}), \projectname{} maintains a low normalized kernel issuing ratio for the inference instance, allowing the collocated training to utilize more SMs. However, MPS-r still keeps a relatively high ratio for inference, resulting in an overall training throughput decrease of 15\% than \projectname{}. Figure~\ref{fig:train-inf-total-kernel-counts} further illustrates total issuing kernel counts in this case, where the purple trace indicates that \projectname{} maintains the highest GPU utilization.
Under the fluctuating workload shown in Figure~\ref{fig:kernel_counts_ratio_inference}, \projectname{} consistently provides more tokens than MPS-r, which explains why the p95 of MPS-r is 3.1$\times$ higher than \projectname{}. 

% Figure~\ref{fig:kernel_counts_ratio_training} is based on <training: LLAMA2-7B, inference: RoBERTa-large> collocation depicted in Figure~\ref{fig:scaling-up/down-train-inf}. 

\textbf{Negligible vertical scaling overhead.} Figure~\ref{fig:train-breakdown} demonstrates that the average throughput loss for different training models with vertical scaling is below 1\%. Similarly, Figure~\ref{fig:inf-breakdown} reveals negligible overhead in inference performance, even as the number of managed instances on a single GPU increases.

\textbf{Sensitivity analysis}. Figure~\ref{fig:sensitivity-maxtokens} shows the impact of the $MaxToken$ size in RCKM on DL performance. A conservative setting severely affects the collocated instances, while an excessively high setting causes interference, particularly in inference tasks.
% It is observed that when the value is set conservatively, collocated instances are severely impacted. When setting too high, it leads to interferes with each other, particularly in inference tasks.

% We set the MaxToken value to 500 million in our experiments.

% \subsubsection{Horizontal Scaling.}
\subsection{Co-Scaling Performance}

\begin{figure}[t]
    \centering
    \includegraphics[scale=0.55]{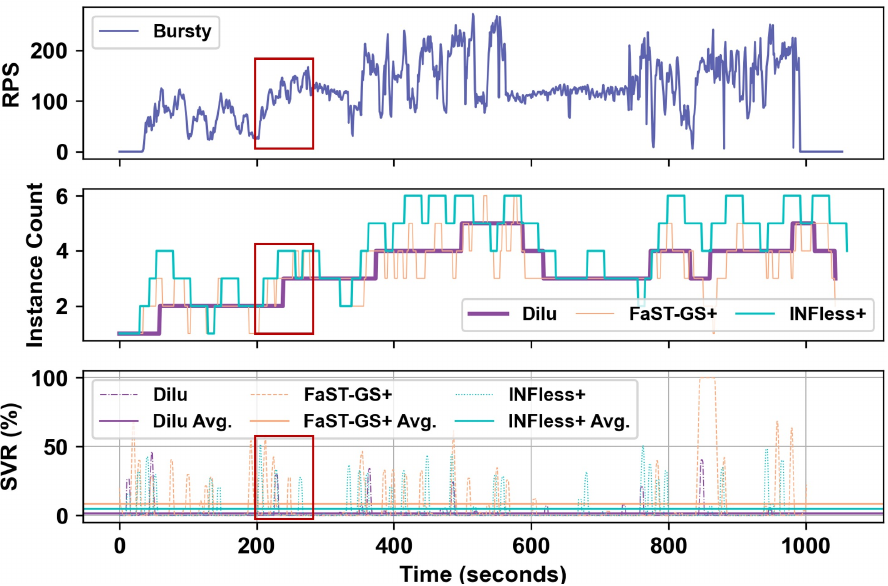}
    % \captionsetup{skip=-0.1pt}
    \caption{Trace analysis on co-scaling performance. SVR denotes the SLO violation rate.}
    \label{fig:scalinginout-combined}
    % \vspace{-0.05in}
\end{figure}

\begin{table}[t]
\caption{Horizontal scaling performance. CSC represents cold start counts. SVR
denotes SLO violation rate. SGT means saved GPU time.}  
\centering

\begin{tabular}{@{}ccccc@{}}
\toprule
\textbf{Trace} & \textbf{Baseline} & \textbf{CSC} & \textbf{SVR(\%)} & \textbf{SGT} \\
\midrule
\multirow{3}{*}{Bursty}  
   & FaST-GS+ & 40 & 10.79 & 715.4s \\
  & INFless+ & 27 & 6.28  & 433.6s \\
  & Dilu    & \textbf{7}  & \textbf{1.79}  & -      \\
\midrule
\multirow{3}{*}{Periodic}  
   & FaST-GS+ & 41 & 19.25 & 650.4s \\
  & INFless+ & 27 & 11.09 & 346.8s \\
  & Dilu    & \textbf{11} & \textbf{9.85}  & -      \\
\midrule
\multirow{3}{*}{Sporadic}  
   & FaST-GS+ & 4  & 7.57  & 65.0s  \\
  & INFless+ & 11 & 5.17  & 216.8s \\
  & Dilu    & \textbf{1}  & \textbf{2.33}  & -      \\
\bottomrule
\end{tabular}
% \vspace{-0.05in}
\label{tab:scalinginout}
\end{table}

% TODO lvcunchi transition
\textbf{Low SLO violation rate and few cold starts}. As shown in Table~\ref{tab:scalinginout}, \projectname{} stands out with the co-scaling support, achieving the lowest SVR and the fewest CSC, while significantly saving GPU costs and ensuring QoS.
% To further explore, we compare \projectname{} with Fast-GS+ and INFless+ of serving RoBERTa-large inferences with typical bursty, sporadic, and periodic workloads.
% To verify with higher elasticity, a classic model RoBERTa-large, which consumes less compute and memory resources, is adopted. 
% Leveraging the fast scaling up \& lazy scaling out strategy, discussed in Section~\ref{sec:lazyscalinginout},
Specifically, \projectname{} achieves an average SVR of 4.7\%, reducing CSC by 75\%-77\% and SVR by 46\%-67\%, compared to INFless+ and FaST-GS+, respectively. FaST-GS+'s eager scaling-out strategy and static scaling-up capability result in a high SVR. INFless+ slightly reduces CSC using prior knowledge but maintains a group of keep-alive instances, leading to substantial GPU waste.

\textbf{Smooth transition.} The co-scaling mechanism brings a smooth transition to guarantee inference SLOs. To investigate further, we record and analyze the serving trace under bursty workloads, as shown in Figure~\ref{fig:scalinginout-combined}.

At certain periods, e.g., framed 200-240 seconds, there exists a workload surge as the top subfigure shows. The fast scaling-up capability provided by RCKM ensures sufficient resource provisioning for these large number of instantaneous requests, thereby securing enough response time to scale out new instances, as indicated by the increase of instance count around 225 seconds in the medium subfigure.

\textbf{Low horizontal scaling overhead.} Regarding the scheduling overhead, \projectname{} generates scheduling decisions for 3,200 instances concurrently within 1.12 seconds. For real-world workloads, the scaling overhead of each instance is less than 1 ms.

% \begin{table}[ht]
% \centering
% \label{tab:performance_comparison}
% \begin{tabular}{@{}ccccc@{}}
% \toprule
% \textbf{Trace} & \textbf{Baseline} & \textbf{CSC} & \textbf{SVR(\%)} & \textbf{SGT} \\
% \midrule
% \multirow{}{}
%          & FaST-GS & 40 & 10.79 & 715.4s \\
%   Bursty & INFless & 27 & 6.28  & 433.6s \\
%          & Dilu    & \textbf{7}  & \textbf{1.79}  & -      \\
% \midrule
% \multirow{}{}
%           & FaST-GS & 41 & 19.25 & 650.4s \\
% Periodic  & INFless & 27 & 11.09 & 346.8s \\
%           & Dilu    & \textbf{11} & \textbf{9.85}  & -      \\
% \midrule
% \multirow{}{}
%          & FaST-GS & 4  & 7.57  & 65.0s  \\
% Sporadic & INFless & 11 & 5.17  & 216.8s \\
%          & Dilu    & \textbf{1}  & \textbf{2.33}  & -      \\
% \bottomrule
% \caption{Scaling in/out Performance TODO explain the metrics}
% \label{tab:scalinginout}
% \end{tabular}
% \end{table}

% \subsubsection{Workload-affinitive Scheduling}
\subsection{End-to-End Scheduling Performance}

\begin{figure}
    \centering
    \includegraphics[scale=0.43]{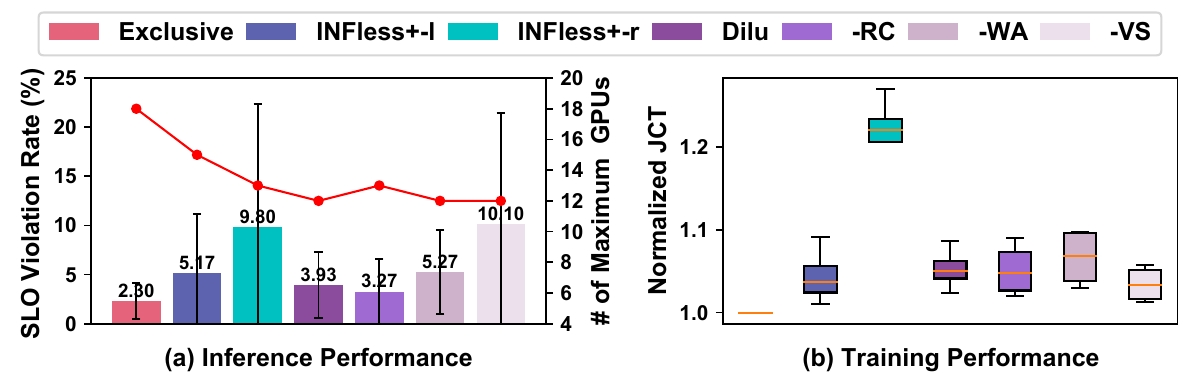}
    % \captionsetup{skip=-0pt}
    \caption{End-to-end performance comparison and component analysis of \projectname{} in local cluster. The error bar in (a) represents the standard deviation of all inference function's SVRs and each boxplot of (b) is constructed based on all normalized training JCT relative to the Exclusive. -RC: without resource complementarity and multi-GPU LLM deployment; -WA: without workload affinity; -VS: without vertical scaling.}
    \label{fig:localcluster}
    % \vspace{-0.05in}
\end{figure}

\begin{figure}
    \centering
    \includegraphics[scale=0.42]{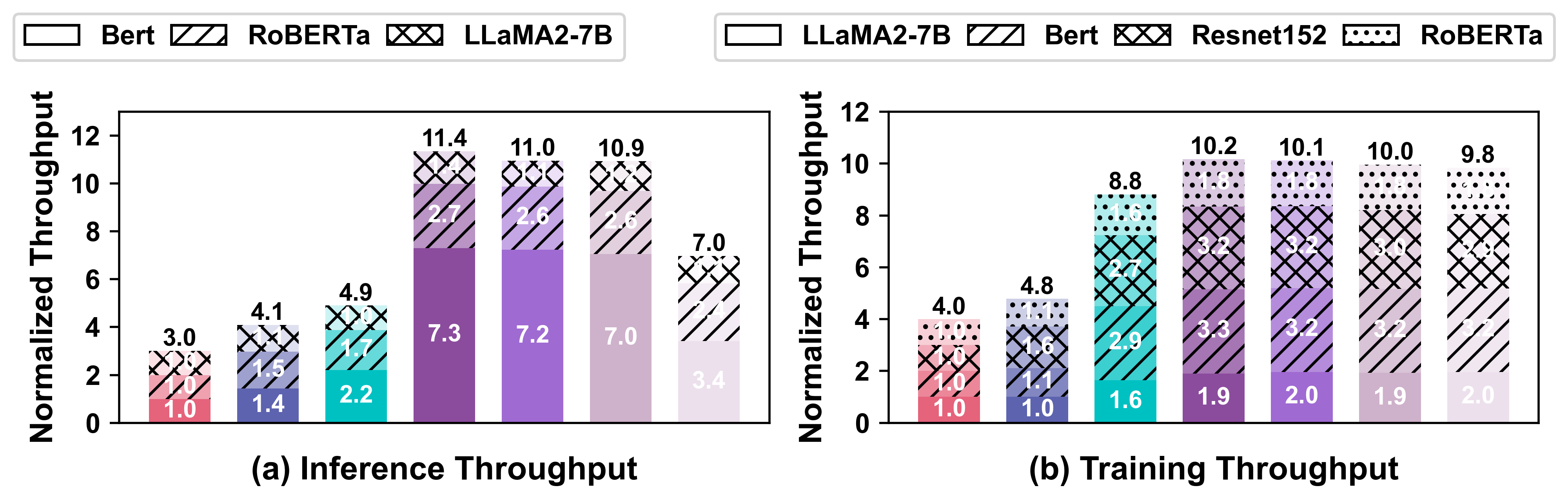}
    % \captionsetup{skip=-0pt}
    \caption{Aggregate throughput performance comparison.}
    \label{fig:localcluster-throughput}
    % \vspace{-0.05in}
\end{figure}

\textbf{Few resource fragments and high throughput}. 
To demonstrate the effectiveness of our scheduling mechanism, we submit 4 training functions at different times, including 2 with 2-workers and 2 with 4-workers, along with three inference functions with varying workloads (specifically, bursty, periodic, and Poisson distributions). 
% Besides training JCT and inference SVR, we also evaluate the maximum number of GPUs (red line) occupied over the period (i.e., 25 minutes).
Figure~\ref{fig:localcluster} shows the end-to-end results.
Although Exclusive achieves the best DL performance, it requires 1.5$\times$ GPUs compared to \projectname{}. INFless+-r performs poorly due to its lower-bound resource allocation. INFless+-l achieves comparable training performance to Dilu but occupies three more GPUs. 
Figure~\ref{fig:localcluster-throughput} reveals that \projectname{} obtains the highest aggregate throughput (i.e., the inference RPS or training throughput divided by the resources they occupy, following a similar definition as INFless \cite{infless}). Specifically, it achieves 3.8$\times$, 2.8$\times$, and 2.3$\times$  the performance of Exclusive, INFless+-l, and INFless+-r in inference, and 2.5$\times$, 2.1$\times$, and 1.2$\times$ in training.

% Specifically, it surpasses Exclusive, INFless+-l, and INFless+-r by 3.8$\times$, 2.8$\times$, and 2.3$\times$ in inference, and 2.5$\times$, 2.1$\times$, and 1.2$\times$ in training. 

% Notably, the -VS suffers a significant 38\% decrease in inference throughput due to severe contention.

\textbf{Ablation study.} Without \textit{RC}, though there is a slight gain in SVR, it requires one additional GPU due to the lack of a distributed deployment strategy for LLM instances. Without \textit{WA}, a slight decline in both training and inference performance occurs due to the barrel effect. Without \textit{VS}, as training infringes on more compute resources, a slight reduction in overall training Job Completion Time (JCT) is observed. However, the average and maximum inference SVR increase by 158\% and 203\%, respectively, compared to \projectname{}. 

% \textbf{Negligible Overhead.} Regarding the scheduling overhead, \projectname{} can generate scheduling decisions for 3,200 instances concurrently within 1.12 seconds. Considering the real-world scheduling workload, the scaling overhead of each instance is less than 1 ms.

% \begin{figure}
%     \centering
%     \subfigure[Inference throughput]{
%         \includegraphics[scale=0.405]{figures/inference_throughput.pdf}
%         \label{fig:inference-throughput}
%     }
%     \hspace{-0.05\linewidth} % Adjust spacing between subfigures
%     \subfigure[Training throughput]{
%         \includegraphics[scale=0.405]{figures/training_throughput.pdf}
%         \label{fig:training-throughput}
%     }
%     \captionsetup{skip=-0.1pt}
%     \caption{Throughput comparison.}
%     \label{fig:localcluster-throughput}
% \end{figure}

\subsection{Large Scale Cluster Simulation}\label{subsec:simu}
\label{sec:large-scale-simulation}

\begin{figure}
    \centering
    \includegraphics[scale=0.57]{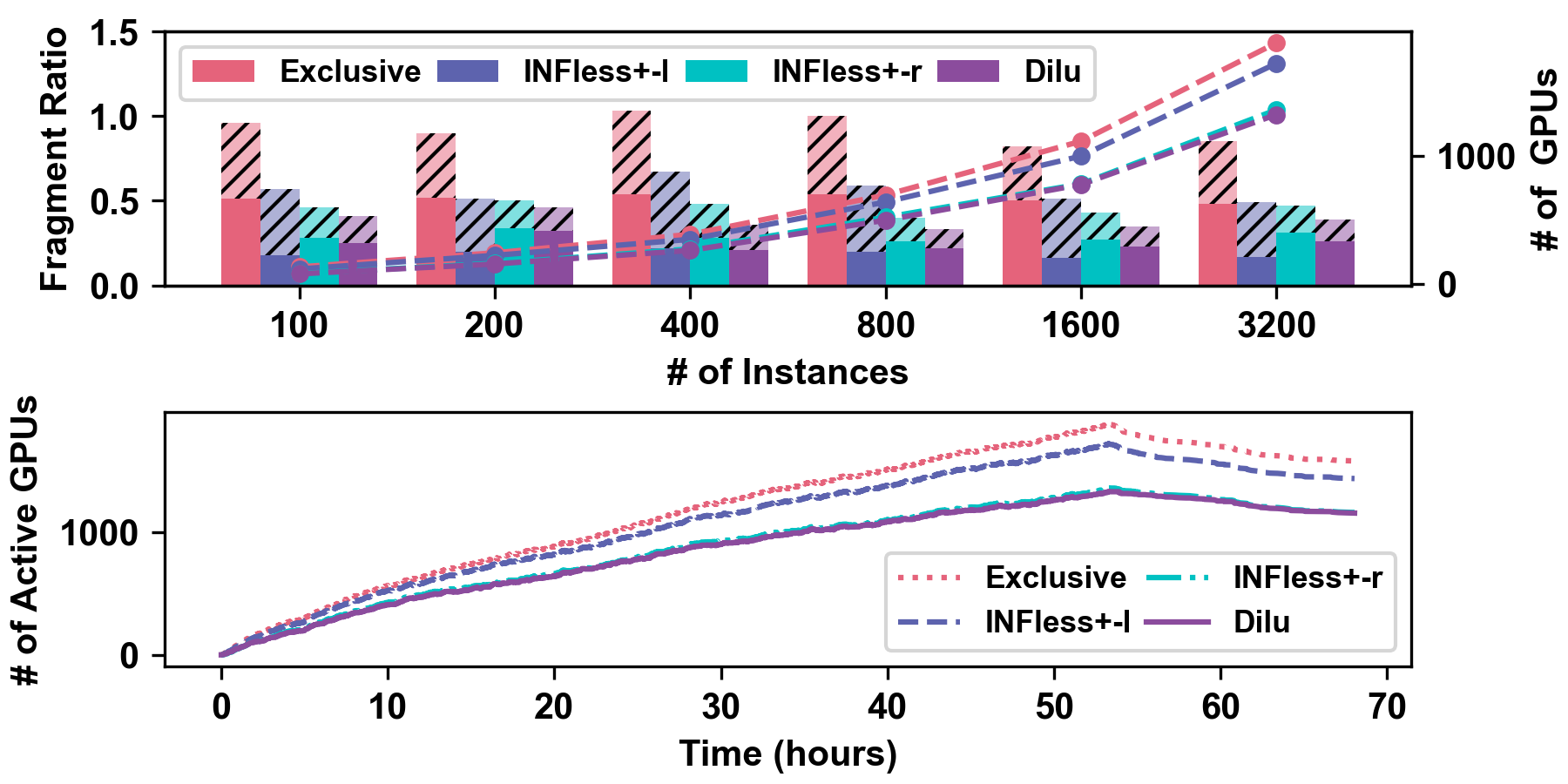}
    % \captionsetup{skip=-0.1pt}
    \caption{GPU provisioning efficiency in large-scale simulations. The dark bottom and light-striped bars represent SM and memory fragments respectively.}
    \label{fig:largescalesimulations}
    % \vspace{-0.15in}
\end{figure}

\textbf{Minimal GPU occupancy and least resource fragmentation}. The results in Figure~\ref{fig:largescalesimulations} indicate that \projectname{} maintains the lowest memory and SM fragmentation, thereby minimizing GPU occupancy. Compared to Exclusive and INFless+-l, Dilu reduces costs by 30\% and 23\% respectively, at a scale of 3,200 instances. The bottom of Figure~\ref{fig:largescalesimulations} presents the variation in GPU count over time, corresponding to the dynamic launching and termination of instances. The consistently lower purple line indicates that \projectname{} keeps the most efficient GPU occupying. Meanwhile, Exclusive and INFless+-l exhibit an increasing gap in GPU usage compared to \projectname{} as the number of instances grows.
% Infless-r, due to its conservative resource provisioning, utilizes many GPUs comparable to that of Dilu while a little higher than  \projectname{} lacking cross-GPU deployment.

% \textbf{Low horizontal scaling overhead.} Regarding the scheduling overhead, \projectname{} generates scheduling decisions for 3,200 instances concurrently within 1.12 seconds. For real-world workloads, the scaling overhead of each instance is less than 1 ms.

\begin{figure}
    \centering
    \subfigure[Oversubscription coefficient]{
        \includegraphics[scale=0.51]{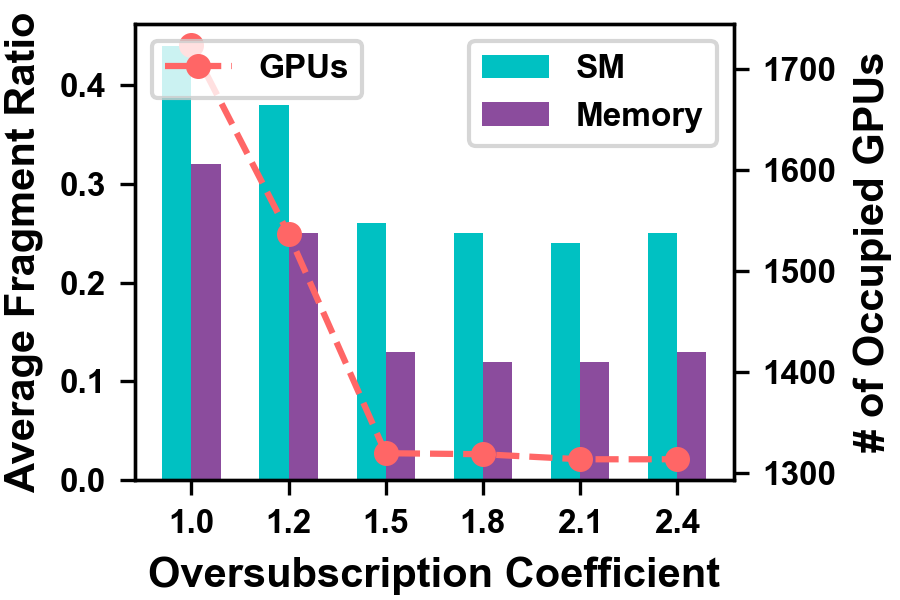}
        \label{fig:sensitivity-oversubmit}
    }
    % \hspace{-0.07\linewidth} % Adjust spacing 
    \subfigure[Max tokens]{
        \includegraphics[scale=0.51]{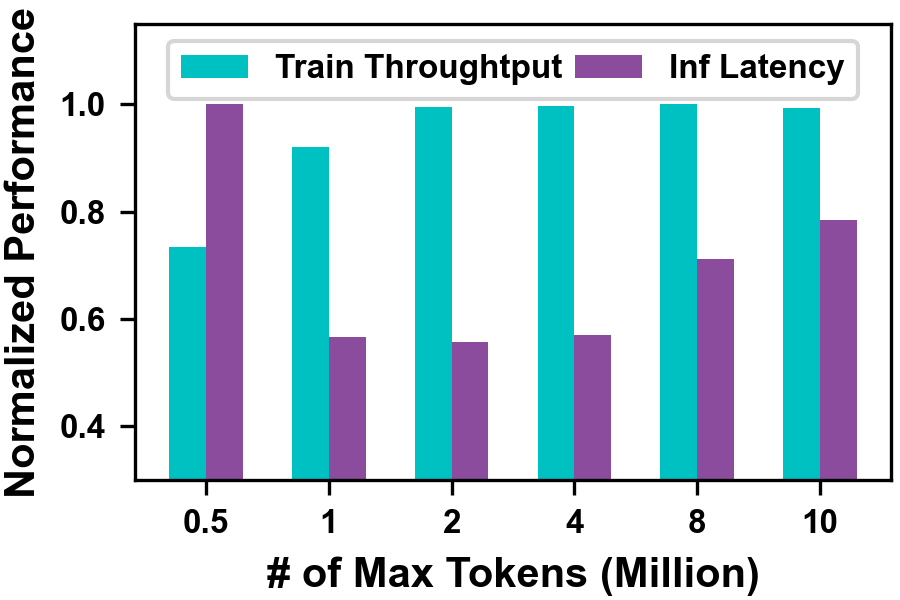}
        \label{fig:sensitivity-maxtokens}
    }
    \captionsetup{skip=-0.1pt}
    \caption{Sensitivity analysis on oversubscription coefficient and max tokens. The oversubscription coefficient denotes the sum of \textit{limit} for collocated instances.}
    \label{fig:Sensitivity}
    % \vspace{-0.15in}
\end{figure}

\textbf{Sensitivity analysis.} Figure~\ref{fig:sensitivity-oversubmit} is based on the 3,200 instances. As the oversubscription coefficient increases, resource fragments and GPU occupancy gradually decrease, with diminishing returns beyond 1.5. Since excessive oversubscription can degrade QoS, we set this parameter to 1.5 in our experiments.

% \subsection{Sensitivity Analysis}
% In this section, we analyze the impact of the oversubscription coefficient (limits of collocated instances on a single GPU) and RCKM's MaxToken on system performance. 

\section{Related Work}

% \textbf{Serverless DL Systems.} Studies propose constructing serverless systems to ease deployment and save costs for both DL training and inference tasks. Studies like LambdaML\cite{lambdaml}, Siren\cite{siren} and Cirrus\cite{cirrus} employ data-parallelism to train models, while FuncPipe\cite{funcpipe} and Hydrozoa\cite{hydrozoa} support model- and pipeline-parallelism. $\lambda$DNN\cite{lambdadnn} dynamically adjusts training instances to accelerate convergence. ElasticFlow\cite{elasticflow} changes training worker numbers and topology according to the scaling curve, increasing job deadline satisfaction rate. Regarding inference,  INFless\cite{infless} fully utilizes the horizontal elasticity of serverless to handle bursty workloads. Amps\cite{amps} and Gillis\cite{gillis} enable model-parallelism in Serverless inference. MArk\cite{mark2019} and Tetris\cite{tetris} separately focus on optimizing CPU and host memory usage to save cost.  BATCH\cite{batch,vldb_inf} and GPUlet\cite{choi2022serving} leverages batching to enhance inference throughput. Furthermore, DGSF\cite{dgsf} and \cite{infless,fastg,mlsyssla} integrate GPUs to guarantee serving SLOs. ServerlessLLM\cite{fu2024serverlessllm} leverages memory-locality to decrease LLM-instance launching time.  However, these studies either monopolize GPUs or statically allocate GPU resources to DL instances, leading to significant spatio-temporal fragmentation. Additionally, the frequent and neglected cold starts of DL instances exacerbate overheads.
\textbf{Serverless DL Systems.} In general, studies aim to simplify deployment and reduce costs for both DL training and inference tasks through the classic horizontal elasticity of serverless computing. LambdaML \cite{lambdaml}, Siren \cite{siren} and Cirrus \cite{cirrus} build data-parallel workers with serverless functions, while FuncPipe \cite{funcpipe} and Hydrozoa \cite{hydrozoa} extend it to the hybrid-parallelism pattern. Works like $\lambda$DNN \cite{lambdadnn} and ElasticFlow \cite{elasticflow} explore to leverage serverless elasticity to accelerate training.
% ElasticFlow \cite{elasticflow} changes training  worker numbers and topology according to the scaling curve, increasing job deadline satisfaction rate.
For inference, studies such as INFless \cite{infless,fastg} focus on improving throughput. Amps \cite{amps} and Gillis \cite{gillis} enable distributed inference in serverless context. MArk \cite{mark2019} and Tetris \cite{tetris} enhance CPU and host memory utilization separately to optimize resource usage. ServerlessLLM \cite{fu2024serverlessllm} leverages memory locality to decrease the launching time of LLM-inference functions. 
% DGSF \cite{dgsf} and other works \cite{infless,fastg,mlsyssla} integrate GPUs to serverless systems to accelerate inference.
 % BATCH \cite{batch} introduces batching to serverless inference functions to acquire higher throughput.
% However, the studies either monopolize GPUs or statically allocate GPU resources to DL instances, leading to significant spatio-temporal fragmentation. Additionally, the frequent and neglected cold starts of DL instances exacerbate overheads.
% However, the studies either monopolize GPUs or statically allocate GPU resources to DL instances, leading to significant spatio-temporal fragmentation. Additionally, the frequent and neglected cold starts of DL instances exacerbate overheads.
Considering the granularity of serverless elasticity, these studies inevitably produce GPU resource fragments due to the adoption of static GPU allocation methods and classic horizontal scaling patterns. In contrast, \projectname{} enables GPU resourcing-on-demand for serverless DL serving, leveraging fine-grained and introspective elasticity.

\textbf{GPU Sharing}. 
% Studies are mainly divided into spatial and temporal methods in GPU sharing to improve GPU utilization. NVIDIA MIG \cite{mig} supports physical partition, and NVIDIA MPS \cite{MPS} provides logical partition by limiting the active CUDA threads percentage of each process, used by  \cite{infless,fastg,dhakal2020gslice,choi2022serving}. Orion \cite{orion} designs a kernel-scheduling dispatching mechanism to share GPU but in thread-level, not adapted to cloud containers. As for temporal sharing, Nexus \cite{shen2019nexus} introduces a squishy bin-packing algorithm to time-share two tasks. TGS \cite{tgs}, Antman \cite{antman} and Baymax \cite{chen2016baymax}prioritize more timeshare for productive jobs, while these only execute one job at a time.  However, Dilu aims to maximize GPU utilization via dynamic GPU limits rate while ensuring each DL task QoS via lower-bounded GPU requests rate, achieving efficient resource-on-demand GPU provisioning.
Previous works attempt to enable GPU sharing in either spatial or temporal dimensions. NVIDIA MIG \cite{NVIDIAMIG} supports physical partitioning of GPU compute resources, and MPS \cite{NVIDIAMPS} offers logical partitioning by limiting the active CUDA threads percentage per process, as adopted by  \cite{infless,fastg,dhakal2020gslice,choi2022serving}. Orion \cite{orion} designs a kernel-scheduling mechanism to share GPU with local threads, but not suitable in the cloud. TGS \cite{tgs} and Antman \cite{antman} prioritize productive jobs to occupy GPUs in timeshare. GPUlet \cite{choi2022serving} and FaST-GS \cite{fastg} enable spatio-temporal sharing to improve inference throughput, while depending on static MPS and are hard to adjust quickly in serverless context.  \projectname{} introduces dynamic GPU provisioning through \textit{<request, limit>} quota control and high-level scheduling to support spatio-temporal sharing.

% \projectname{} introduces dynamic GPU provisioning via \textit{<request, limit>} quotas and high-level scheduling to support spatio-temporal sharing.
% Nexus \cite{shen2019nexus} introduces a squishy bin-packing algorithm to enable two-task GPU timeshare. 

\textbf{DL Scheduling}. 
With the rise of deep learning, many studies focus on scheduling DL tasks to improve resource utilization. Antman \cite{antman}, Tiresias \cite{gu2019tiresias} and Pollux \cite{qiao2021pollux} concentrate on reducing the average JCT for training tasks while ElasticFlow \cite{elasticflow} and Chronus \cite{gao2021chronus} are deadline-aware. For inference workloads, although INFless \cite{infless} and FaST-GS \cite{fastg} optimize inference throughput and guarantee QoS through existing horizontal-scaling mechanisms, they fail to fully utilize temporal resource fragmentation. Moreover, these works cannot adjust GPU provisioning at the 5ms granularity, whereas Dilu can.

\section{Conclusion and Discussion}
Given the rising trends in serverless DL serving and apparent inefficiencies in current GPU provisioning methods, we present \projectname{}, a cross-layer and introspective design to improve GPU utilization and extend the elastic scaling dimensions of serverless DL systems. The adaptive 2D co-scaling mechanism not only dynamically adjusts GPU provisioning at the intra-instance level to minimize fragmentation, but also enhances the capability to handle sudden workloads, and reduces cold starts at the inter-instance level while guaranteeing QoS. To our knowledge, \projectname{} is the first serverless DL system with dynamic and omultidimensional elasticity for heterogeneous DL functions.

In the future, we plan to further extend \projectname{} to explore more elastic serverless training and LLM serving. Additionally, GPU-sharing protection will be aligned with low-level security mechanisms.

\section*{Acknowledgement}
We thank all anonymous reviewers for their valuable feedback and comments. We also thank our colleague Xiaohong Wang for her kind support in this study. This work is supported by the Innovation Funding of ICT, CAS under Grant No. E361060, No. E461040, and the Pilot for Major Scientific Research Facility of Jiangsu Province of China under Grant No. BM2021800.

\normalem
\bibliographystyle{plain}
\balance
\bibliography{main}
\end{document}